\documentclass[11pt]{article}

\usepackage[dvips]{graphicx}
\usepackage{psfrag}
\usepackage[english]{babel}
\usepackage{cite}

\usepackage{color}

\usepackage{amsmath}

\allowdisplaybreaks[2]

\newcommand{\re}{\operatorname{Re}}
\newcommand{\im}{\operatorname{Im}}

\newcommand{\gev}{\operatorname{GeV}}
\newcommand{\mev}{\operatorname{MeV}}

\newcommand{\fb}{\operatorname{fb}}

\newcommand{\ms}{\mskip 1.5mu}

\newcommand{\msp}{\phantom{-}}

\newcommand{\bm}[1]{\boldsymbol #1}

\newcommand{\gsim}{\raisebox{-4pt}{%
    $\,\stackrel{\textstyle >}{\sim}\,$}}

\newcommand{\tmax}{|t|_{\text{max}}}
\newcommand{\xb}{x_{\!B}}
\newcommand{\mn}{m_{\!N}}

\begin{document}

\begin{flushright}
DESY 08-103 \\
CPHT-RR007.0308
\end{flushright}

\begin{center}
\vspace{4.0\baselineskip}
{\LARGE \bf Deeply virtual Compton scattering \\[0.2\baselineskip]
  on a virtual pion target} \\
\vspace{4.0\baselineskip}
{\large D.~Amrath$^{\,a}$, M.~Diehl$^{\,a}$ and J.~P.~Lansberg$^{\,b,c}$}
\\[\baselineskip]
\end{center}

\noindent\textit{%
$^a$ Deutsches Elektronen-Synchroton DESY, 22603~Hamburg,
  Germany \\
$^b$ Centre de Physique Th\'eorique, \'Ecole Polytechnique, CNRS,
91128 Palaiseau, France \\
$^c$ Institut f\"ur Theoretische Physik, Universit\"at
Heidelberg, Philosophenweg 19, 69120~Heidelberg, \\
\phantom{$^c$} Germany}
\vspace{4.0\baselineskip}

\begin{center}
\textbf{{\large Abstract}}\\[0.5\baselineskip]
\parbox{0.9\textwidth}{We study deeply virtual Compton scattering on a
  virtual pion that is emitted by a proton.  Using a range of models for
  the generalized parton distributions of the pion, we evaluate the cross
  section, as well as the beam spin and beam charge asymmetries in the
  leading-twist approximation.  Studying Compton scattering on the pion in
  suitable kinematics puts high demands on both beam energy and
  luminosity, and we find that the corresponding requirements will first
  be met after the energy upgrade at Jefferson Laboratory.  As a
  by-product of our study, we construct a parameterization of pion
  generalized parton distributions that has a non-trivial interplay
  between the $x$ and $t$ dependence and is in good agreement with form
  factor data and lattice calculations.}
\end{center}

\newpage

%%%%%%%%%%%%%%%%%%%%%%%%%%%%%%%%%%%%%%%%%%%%%%%%%%%%%%%%%%%%%

\section{Introduction}
\label{sec:intro}

The concept of generalized parton distributions (GPDs) is a versatile tool
to describe hadron structure at the quark-gluon level and has given rise
to vigorous theoretical and experimental activities.  Among the attractive
features of GPDs are the possibilities to connect ordinary parton
densities with elastic form factors \cite{Ji:1998pc} and to explore the
spatial distributions of partons inside a hadron
\cite{Burkardt:2002hr,Ralston:2001xs}.  Reviews of this extensive subject
can be found in
\cite{Goeke:2001tz,Diehl:2003ny,Belitsky:2005qn,Boffi:2007yc}.

The pion plays a special role in the low-energy sector of QCD as the
lightest bound state and the Pseudo-Goldstone boson associated with chiral
symmetry breaking.  Given the difficulty to perform high-energy
experiments with a pion in the initial state, our knowledge of its
internal structure is, however, scarce compared with what is known about
the nucleon.  Currently, the principal sources of information are the
spacelike electromagnetic form factor of the pion
\cite{Amendolia:1986wj,Ackermann:1977rp,Tadevosyan:2007yd,Horn:2006tm} and
its parton densities extracted from Drell-Yan production with pion beams
\cite{Sutton:1991ay,Gluck:1999xe,Wijesooriya:2005ir}.  Measurements
constraining the GPDs of the pion would provide a natural extension of
this knowledge.  On the theoretical side, the pion GPDs have been studied
in a number of dynamical models \cite{Anikin:2000th} and on the lattice
\cite{Brommel:2005ee,Brommel:2007zz,Brommel:2007xd,Dalley:2003sz}.
Important theoretical investigations have been performed for a pion target
in the first instance, because of its relative simplicity as a spin-zero
particle, see for instance \cite{Polyakov:1999gs,Anikin:2000em}.

\begin{figure}
\begin{center}
\includegraphics[width=0.7\textwidth]{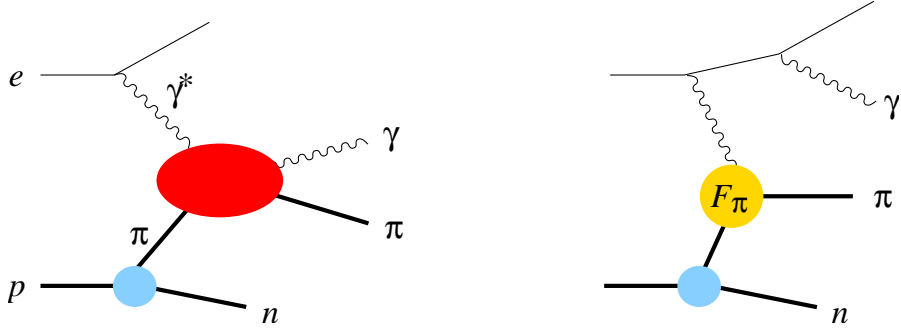}
\caption{\label{fig:graph} Graphs for $ep\to e\gamma\pi^+ n$ in the
  one-pion exchange approximation.  Contributing subprocesses are virtual
  Compton scattering on a pion (left) and the Bethe-Heitler process
  (right).  The crossed Bethe-Heitler graph (not shown) has the photons
  attached to the lepton line in opposite order.  The blob marked with
  $F_\pi$ represents the electromagnetic pion form factor.}
\end{center}
\end{figure}

The purpose of the present work is to estimate how pion GPDs may be
investigated in deeply virtual Compton scattering (DVCS), which is the
theoretically cleanest and most advanced among the hard exclusive
processes involving generalized parton distributions.  We consider the
reaction $ep\to e\gamma\pi^+ n$ at small invariant momentum transfer
between the proton and neutron.  In the one-pion exchange approximation,
the reaction is then described by the emission of a slightly off-shell
pion from the proton, followed by the scattering process $e\pi^+ \to
e\gamma\pi^+$.  This can be seen as an analog of the reaction $ep\to
e\pi^+ n$, which has been used to extract the electromagnetic pion form
factor for all but the smallest values of momentum transfer
\cite{Ackermann:1977rp,Tadevosyan:2007yd,Horn:2006tm}.
Two mechanisms contribute to $e\pi \to e\gamma\pi$, namely virtual Compton
scattering and the Bethe-Heitler process, as shown in
Fig.~\ref{fig:graph}.  Suitable strategies for isolating the Compton
signal are the same as those for scattering on a nucleon target, which
have been elaborated in detail \cite{Diehl:1997bu} and successfully used
in experiment \cite{Airapetian:2001yk,H1:2007ep}.  The corresponding
expressions for the pion can be found in \cite{Belitsky:2000vk}, where
also numerical estimates for $e\pi \to e\gamma\pi$ have been given.
We note that in suitable kinematics, the reaction $ep\to e\gamma\pi^+ n$
may also be used to study virtual Compton scattering on the pion in the
backward region, whose description involves the so-called transition
distribution amplitude from a pion to a photon \cite{Pire:2004ie}.  In
experiments with a real photon beam, the lepton-pair production process
$\gamma p\to e^+e^-\ms \pi^+ n$ can provide access to timelike Compton
scattering $\gamma\pi \to \gamma^*\pi$ on the pion, which is closely
related to DVCS by crossing \cite{Berger:2001xd}.

Our paper is organized as follows.  In the next section we present in some
detail the kinematics of $ep\to e\gamma\pi^+ n$.  In Sect.~\ref{sec:cross}
we give the basic equations for the analysis of this process in the
one-pion exchange picture and in the framework of generalized parton
distributions.  We also briefly discuss the validity of the one-pion
exchange approximation.  In Sect.~\ref{sec:pi-gpd} we present a model for
the GPDs of the pion, paying special attention to their $t$-dependence.
In Sect.~\ref{sec:cross-est} we finally give estimates for cross sections
and asymmetries in the HERMES experiment and at Jefferson Lab, before
summarizing our main findings in Sect.~\ref{sec:sum}.

%%%%%%%%%%%%%%%%%%%%%%%%%%%%%%%%%%%%%%%%%%%%%%%%%%%%%%%%%%%%%
%%%%%%%%%%%%%%%%%%%%%%%%%%%%%%%%%%%%%%%%%%%%%%%%%%%%%%%%%%%%%

\section{Kinematics}
\label{sec:kin}

In this section we discuss the kinematics of the reaction
\begin{equation}
  \label{proc}
e(l) + p(p) \to e(l') + \gamma(q') + \pi^+(p'_\pi) + n(p') \,,
\end{equation}
with four-momenta given in parentheses.  We write
\begin{align}
q     &= l-l' \,, &
p_\pi &= p-p'
\end{align}
for the four-momenta of the virtual photon and the virtual pion, and use
the standard variables
\begin{align}
Q^2 &= -q^2 \,, &  W^2 &= (p+q)^2 \,, &  s &= (p+l)^2 \,, &
\xb &= \frac{Q^2}{2p\cdot q} \,, &
y   &= \frac{p\cdot q}{p\cdot l}
\end{align}
for deep inelastic scattering processes.  The variables
\begin{align}
t     &= (p-p')^2 \,, &
x_\pi &= \frac{p_\pi \cdot l}{p \cdot l}
\end{align}
describe the emission of the virtual pion from the proton target, where
$x_\pi$ is the fraction of energy that the virtual pion takes away from
the proton in the $ep$ c.m.  The azimuthal angles of the final-state
electron and neutron in that frame are respectively denoted by $\psi_e$
and $\psi_n$, where the $z$-axis is chosen along the lepton beam momentum.
We write $\mn$ and $m_\pi$ for the nucleon and pion masses, and neglect
the lepton mass throughout our work.
An important role is played by the kinematic limit
\begin{equation}
  \label{t-min}
-t \ge -t_0 = \frac{x_\pi^2 \mn^2}{1-x_\pi} \,,
\end{equation}
which is readily obtained from the expression
$\bm{p}'^2_T = -t\ms (1-x_\pi^{}) - x_\pi^2 \mn^2$
for the squared transverse momentum of the neutron in the $ep$ c.m.  We
need two more variables to describe the $\gamma$ and $\pi$ in the final
state, namely
\begin{equation}
t_\pi^{} = (p^{}_\pi - p'_\pi)^2
\end{equation}
and the azimuthal angle $\phi_\pi$ between the plane spanned by
$\bm{p}^{}_\pi$ and $\bm{p}'_\pi$ and the plane spanned by $\bm{l}$ and
$\bm{l}'$ in the c.m.\ of $\pi\gamma$ in the final state.  For the sign of
$\phi_\pi$ we follow the usual convention for DVCS on a proton
target.\footnote{%
  One thus obtains $\phi_\pi$ from the angle $\phi_h$ in
  \protect\cite{Bacchetta:2004jz} by replacing $P\to p_\pi$ and $P_h\to
  p'_\pi$.}
For our later discussion it is useful to introduce further variables,
which refer to the subprocess $e\pi \to e\gamma\pi$ on the virtual pion
target, namely
\begin{align}
s_\pi   &= (p_\pi + q)^2 \,, &
\xb^\pi &= \frac{Q^2}{2 p_\pi \cdot q} \,, &
y_\pi   &= \frac{p_\pi \cdot q}{p_\pi \cdot l} \,.
\end{align}
One finds
\begin{align}
  \label{s-pi}
s_\pi &= Q^2\ms \Bigl( \frac{x_\pi}{\xb} - 1 \Bigr)
  - 2 \cos(\psi_e-\psi_n)\, \sqrt{(1-y) Q^2 - (y \xb\ms \mn)^2\ms }\,
      \sqrt{(1-x_\pi) (t_0-t)}
\nonumber \\
 &\quad + 2 y \xb\ms x_\pi \mn^2 + (1-y \xb)\ms t \,.
\end{align}

To select kinematics where DVCS on a pion can be described in terms of
generalized parton distributions, we take the Bjorken limit
\begin{align}
Q^2 &\to \infty & \text{at fixed $y$, $\xb$, $x_\pi$, $t$, $t_\pi$.}
\end{align}
The squared c.m.\ energies $s$, $W^2$, and $s_\pi$ then become large,
whereas the squared momentum transfers $t$ and $t_\pi$ are small compared
with $Q^2$.  The relation \eqref{s-pi} then implies
\begin{align}
  \label{xBpi}
\xb^\pi    &\approx \frac{\xb}{x_\pi} \,, &
y_\pi      &= y \frac{\xb}{ x_\pi\ms \xb^\pi} \approx y
\end{align}
and
\begin{equation}
  \label{xpi-y}
x_\pi\ms y \approx \frac{s_\pi+Q^2}{s}
\end{equation}
with corrections of order $m/Q$, where $m^2$ represents the small scales
$\mn^2$, $m_\pi^2$, $t$ and $t_\pi$, whose magnitudes we do not
distinguish at this point.
The phase space element of the process \eqref{proc} can be written as
\begin{equation}
  \label{phase-space}
\frac{d^3 l'}{2l'^0}\, \frac{d^3 q'}{2q'^0}\, 
\frac{d^3 p'_\pi}{2p'^0_\pi}\, \frac{d^3 p'}{2p'^0}\,
\delta^{(4)}(l'+q'+p'_\pi+p' - l-p)
 = \frac{dQ^2\, dy\, d\psi_e\, dt_\pi\,
     d\phi_\pi\, dt\, dx_\pi\, d\psi_n}{64\ms \sqrt{(s_\pi+Q^2+t)^2 -
     4s_\pi\ms t \rule{0pt}{0.85em}}} \,.
\end{equation}

The interpretation of the process \eqref{proc} in terms of DVCS on a
virtual pion target puts several conditions on the kinematics, which we
now discuss.  First of all we impose an upper cutoff on $|t|$,
\begin{equation}
  \label{t-max}
|t| \le \tmax \,,
\end{equation}
to ensure that the $p\to n$ transition is dominated by virtual pion
emission.  Since $\tmax$ must be bigger than $-t_0$ in \eqref{t-min}, this
implies a maximum value for $x_\pi$,
\begin{align}
  \label{xpi-max}
x_\pi &\le x_{\pi\ms\text{max}}
       = \frac{1}{2} \Bigl[ \sqrt{\tau^2 + 4\tau} - \tau \Bigr]
&
\text{with}~ \tau &= \frac{\tmax}{\mn^2} \,.
\end{align}
Since we want the subprocess $\gamma^*\pi \to \gamma\pi$ to be in Bjorken
kinematics, we further impose lower cutoffs \begin{align}
  \label{s-min}
s_\pi &\ge s_{\pi\ms\text{min}} \,, & Q^2 \ge Q_{\text{min}}^2 \,.
\end{align}
According to \eqref{xpi-y} this implies lower limits on $x_\pi$ and on
$y$,
\begin{align}
  \label{xpi-y-min}
x_{\pi\ms\text{min}} &\approx \frac{1}{y_{\text{max}}} \,
  \frac{s^{}_{\pi\ms\text{min}} + Q_{\text{min}}^2}{s} \,,
&
y_{\text{min}} &\approx \frac{1}{x_{\pi\ms\text{max}}} \,
  \frac{s^{}_{\pi\ms\text{min}} + Q_{\text{min}}^2}{s} \,,
\end{align}
where $y_{\text{max}}$ is an upper limit on $y$ we will later impose both
for theoretical and for experimental reasons (see
Sect.~\ref{sec:cross-est}). The relation \eqref{xpi-y} also restricts the
largest possible values of $Q^2$ to
\begin{equation}
  \label{Q-max}
Q_{\text{max}}^2 \approx
  s x_{\pi\ms\text{max}}^{}\, y_{\text{max}}^{}
  - s_{\pi\ms\text{min}}^{} \,.
\end{equation}
For a clean physical interpretation of the reaction \eqref{proc} as DVCS
on a weakly off-shell pion target, it is desirable to take rather small
$\tmax$ and rather large $s^{}_{\pi\ms\text{min}}$ and $Q_{\text{min}}^2$.
With \eqref{xpi-max} and the first relation in \eqref{xpi-y-min}, this
only leaves enough phase space for $x_\pi$ if the total c.m.\ energy $s$
is sufficiently large.

%%%%%%%%%%%%%%%%%%%%%%%%%%%%%%%%%%%%%%%%%%%%%%%%%%%%%%%%%%%%%

\section{Calculation of the cross section}
\label{sec:cross}

In this section we give some basic formulae for the process $ep\to
e\gamma\pi n$ in the one-pion exchange approximation and discuss the
validity of this approximation.

%%%%%%%%%%%%%%%%%%%%%%%%%%%%%%%%%

\subsection{The one-pion exchange approximation}
\label{sec:one-pi}

In the one-pion exchange approximation, the amplitudes for $ep\to
e\gamma\pi n$ and for $e\pi\to e\gamma\pi$ are related as
\begin{align}
  \label{piNN}
\mathcal{M}_{ep} &= \bar{u}(p')\gamma_5\ms u(p)\,
   \frac{\sqrt{2}\ms g_{\pi NN}}{m_\pi^2-t}\, F(t) \,
   \mathcal{M}_{e\pi} \,,
& F(t) &=
   \frac{\Lambda^2-m_\pi^2}{\Lambda^2-t}
\end{align}
with the pion-nucleon coupling $g_{\pi NN} = 13.05$ \cite{Stoks:1992ja}.
Here we have
introduced a phenomenological factor $F(t)$ to soften the pion-nucleon
vertex when the pion virtuality $t$ becomes large compared to $m_\pi^2$.
In our calculations we will take $\Lambda = 800\mev$ from
\cite{Koepf:1995yh}.
The $ep$ cross section is then given by
\begin{align}
  \label{x-sect-master}
\frac{d^8\sigma(ep\to e\gamma\pi n)}{dy\, dQ^2\, d\psi_e\, dt_\pi\,
      d\phi_\pi\, dt\, dx_\pi\, d\psi_n}
 &= \frac{1}{128\ms (2\pi)^8\, (s-\mn^2) \sqrt{(s_\pi+Q^2+t)^2 -
     4s_\pi\ms t \rule{0pt}{0.85em}}}\,
\nonumber \\
 &\quad \times
   \bigl[ \sqrt{2}\ms g_{\pi NN}\ms F(t) \bigr]^2\,
   \frac{-t}{(m_\pi^2-t)^2}\,
   \sum_{\text{spins}} |\mathcal{M}_{e\pi}|^2 \,,
\end{align}
where $\sum_{\text{spins}}$ sums over the polarizations of the final-state
electron and photon.  In \eqref{x-sect-master} we have further averaged
over the polarization of the proton target and summed over the
polarization of the outgoing neutron, but kept the lepton beam
polarization fixed.
Defining the cross section
\begin{equation}
  \label{epi-Xsect}
\frac{d^4\sigma(e\pi\to e\gamma\pi)}{dy_\pi\, dQ^2\, dt_\pi\,
      d\phi_\pi}
  = \frac{1}{32\ms (2\pi)^4\, x_\pi (s-\mn^2) \sqrt{(s_\pi+Q^2+t)^2 -
      4s_\pi\ms t \rule{0pt}{0.85em}}}\,
    \sum_{\text{spins}} |\mathcal{M}_{e\pi}|^2
\end{equation}
on a virtual pion target, we have the simple relation
\begin{align}
  \label{ep-to-epi}
\frac{d^6\sigma(ep\to e\gamma\pi n)}{dy\, dQ^2\, dt_\pi\,
      d\phi_\pi\, dt\, dx_\pi}
 = x_\pi\,
   \frac{g_{\pi NN}^2}{8\pi^2}\, \bigl[ F(t) \bigr]^2
   \frac{-t}{(m_\pi^2-t)^2}\;
   \frac{d^4\sigma(e\pi\to e\gamma\pi)}{dy_\pi\, dQ^2\,
     dt_\pi\, d\phi_\pi} \,,
\end{align}
where we have integrated over the angles $\psi_e$ and $\psi_n$.

\begin{figure}
\begin{center}
\psfrag{xx}[c]{$x_\pi$}
\psfrag{zz}[b]{$\Pi(x_\pi, \tmax)$}
\includegraphics[width=0.5\textwidth]{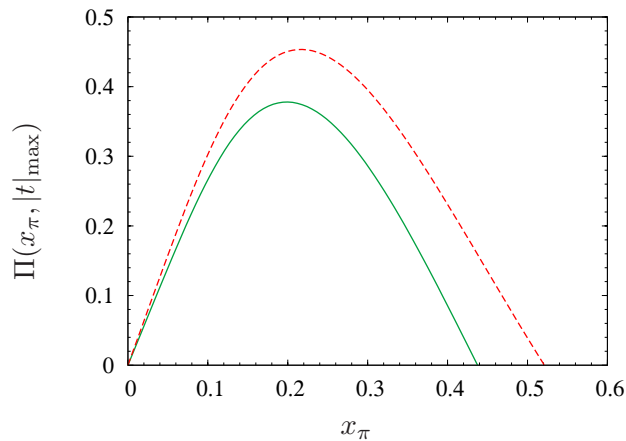}
\caption{\label{fig:pi-flux} The pion flux factor $\Pi(x_\pi, \tmax)$
  defined in \protect\eqref{pi-flux}, shown for $\tmax= 0.3\gev^2$ (solid)
  and $\tmax= 0.5 \gev^2$ (dashed).}
\end{center}
\end{figure}

For a rough estimate one may neglect the dependence of the $e\pi$ cross
section \eqref{epi-Xsect} on the pion virtuality $t$.  Integrating the
factor of proportionality in \eqref{ep-to-epi} over $t$, one then obtains
\begin{equation}
  \label{rough-approx}
\frac{d^4\sigma(ep\to e\gamma\pi n)}{dy\, dQ^2\, dt_\pi\, d\phi_\pi}
\approx \int dx_\pi\, \Pi(x_\pi, \tmax)\;
\frac{d^4\sigma(e\pi\to e\gamma\pi)}{dy_\pi\, dQ^2\, dt_\pi\,
  d\phi_\pi}
\end{equation}
with
\begin{equation}
  \label{pi-flux}
\Pi(x_\pi, \tmax) = x_\pi\, \frac{g_{\pi NN}^2}{8\pi^2}
\int\limits_{-\tmax}^{t_0(x_\pi)} dt\;
   \bigl[ F(t) \bigr]^2
   \frac{-t}{(m_\pi^2-t)^2} \,,
\end{equation}
where $t_0(x_\pi)$ is given in \eqref{t-min}.  Note that the $e\pi$ cross
section on the r.h.s.\ of \eqref{rough-approx} depends on $x_\pi$ via
$s_\pi \approx x_\pi y_\pi\ms s - Q^2$.  With the form factor of the
pion-nucleon vertex taken in \eqref{piNN}, the integral in \eqref{pi-flux}
can be done analytically using
\begin{equation}
\int dt\, \bigl[ F(t) \bigr]^2 \frac{-t}{(m_\pi^2-t)^2}
= \frac{\Lambda^2 + m_\pi^2}{\Lambda^2 - m_\pi^2}\,
    \ln\frac{\Lambda^2 -t}{m_\pi^2 - t}
  - \frac{\Lambda^2}{\Lambda^2 - t}
  - \frac{m_\pi^2}{m_\pi^2 - t} \,.
\end{equation}
In Fig.~\ref{fig:pi-flux} we show the pion flux factor $\Pi(x_\pi, \tmax)$
as a function of $x_\pi$ for two values of $\tmax$.

%%%%%%%%%%%%%%%%%%%%%%%%%%%%%%%%%

\subsection{Validity of the one-pion exchange approximation}
\label{sec:one-pi-disc}

The validity of the one-pion exchange approximation in the process
\eqref{proc} cannot be taken for granted, and it is natural to see what is
known for similar processes.

As already mentioned in the introduction, an important process in this
context is $ep\to e\pi n$, where the one-pion exchange approximation
yields the subprocess $e\pi\to e\pi$, from which the electromagnetic pion
form factor can be extracted.  The most recent measurements
\cite{Tadevosyan:2007yd,Horn:2006tm}, as well as previous data, provide a
clear indication that pion exchange cannot be the only contribution to
this reaction.  This is because the cross section $\sigma_T$ for
$\gamma^*p\to \pi n$ with transverse $\gamma^*$ polarization in the
$\gamma^* p$ c.m.\ is seen to be clearly nonzero, although at low $|t|$ it
is smaller than the cross section $\sigma_L$ for a longitudinal
$\gamma^*$.  To understand the implications of this observation, let us
take the case $|t|=|t|_{\text{min}}$, where by definition the angle
between the proton and neutron momenta is zero in the $\gamma^* p$ c.m.
Due to angular momentum conservation, the subprocess $\gamma^*\pi\to \pi$
can then only proceed for longitudinal photon polarization, i.e., pion
exchange contributes only to $\sigma_L$.  In terms of $t$-channel
exchanges, the presence of $\sigma_T$ is thus an indicator for the
exchange of states with nonzero spin, with the $\rho$ being an obvious
candidate because of its relatively low mass. A corresponding model
calculation in \cite{Guidal:1997hy} indeed yields a nonvanishing
$\sigma_T$, although it undershoots the measured values of
\cite{Tadevosyan:2007yd} and \cite{Horn:2006tm}, which were taken at
$\gamma^* p$ c.m.\ energies of $W= 1.95\gev$ and $W= 2.22\gev$,
respectively.  In \cite{Tadevosyan:2007yd} this mismatch was ascribed to
possible resonance contributions in the $\pi n$ channel.  At
$|t|=|t|_{\text{min}}$ there is no contribution to $\sigma_L$ from $\rho$
exchange because of parity conservation in the subprocess $\gamma^*\rho
\to \pi$, so that for the extraction of the pion form factor a separation
of $\sigma_L$ and $\sigma_T$ should considerably enhance the contribution
from pure pion exchange.

We note that the situation in $\gamma^*p\to \gamma \pi n$ is different.
Even at $|t|=|t|_{\text{min}}$ both subprocesses $\gamma^* \pi\to \gamma
\pi$ and $\gamma^*\rho \to \gamma \pi$ can take place for transverse
$\gamma^*$ polarization, which is dominant in Bjorken kinematics according
to the factorization theorem for DVCS.  Whereas taking low $|t|$ will
enhance pion exchange also in this case, it may be useful to assess the
quantitative importance of $\gamma^*\rho \to \gamma \pi$.  This would
require information about the GPDs for the $\rho\to \pi$ transition.
Simple helicity counting \cite{Diehl:2001pm} shows that at twist-two level
there are two GPDs involving the axial current and one GPD involving the
vector current for the quarks.  Their $x$ moments are accessible to an
evaluation in lattice QCD, which may thus help to estimate the role of
$\rho$ exchange in the reaction $ep\to e \gamma\pi n$.  In particular,
the lowest moment of the vector current GPD gives the electromagnetic
$\rho\to \pi$ transition form factor, which also enters the Bethe-Heitler
process in $e\rho \to e\gamma\pi$.  A further constraint on the $\rho\to
\pi$ transition GPDs can be obtained by taking the limit of soft pion
momentum, in analogy to what has been done for the $N\to \pi$ transition
in \cite{Lansberg:2007ec}.

Concerning resonance contributions, we recall that the $\gamma^* p$ c.m.\
energy $W$ is large in Bjorken kinematics.  In experiments suited to
investigate DVCS, $W$ will thus be much larger than in the pion form
factor measurements \cite{Tadevosyan:2007yd,Horn:2006tm}.  More
problematic are possible resonances in the $\pi n$ channel of the final
state, which are of course not taken into account by the one-pion exchange
approximation.  The relevant invariant mass is
\begin{align}
  \label{snpi}
M^2_{\pi n} &=
(p'_\pi + p')^2 = \mn^2 + 2\mn^{} E_\pi + t_\pi - t
\nonumber \\[0.5em]
 &= \mn^2 + m_\pi^2 + \frac{1}{x_\pi} \biggl[
      (t_{\pi 0} - t_\pi) (1-x_\pi) - t\ms (1-\xb^\pi) +
      \frac{1-x_\pi}{1-\xb^\pi}\, m_\pi^2
\nonumber \\
 &\qquad
    + 2 \cos(\phi_\pi +\psi_e-\psi_n)
      \sqrt{(1-x_\pi^{}) (1-\xb^\pi) (t_0-t) (t_{\pi 0}-t_\pi)}
    \,\biggr]
    + \mathcal{O}\Bigl( \frac{m^3}{Q} \Bigr) \,.
\end{align}
Here $E_\pi$ is the energy of the outgoing pion in the proton rest frame,
and $t_{\pi 0}$ is the upper kinematic limit of $t_\pi$.  In the Bjorken
limit one has
\begin{equation}
  \label{tpi-0}
t_{\pi 0} \,\approx\,
  \xb^\pi\, t - \frac{\xb^\pi\ms m_\pi^2}{1-\xb^\pi}
\end{equation}
with corrections of order $(\xb^\pi\ms m)^2/Q^2$.  In our numerical
calculations we use the exact expression of $t_{\pi 0}$, so that any
$\phi_\pi$ dependent term exactly vanishes at the kinematical point $t_\pi
=t_{\pi 0}$, where $\phi_\pi$ is not defined.
We see from \eqref{snpi} that to avoid resonances contributions in the
$\pi n$ channel it is advantageous to have low $x_\pi$ and relatively
large $|t_\pi|$ (while still respecting the condition $|t_\pi| \ll Q^2$
for Bjorken kinematics).

A different process relevant in our context is deep inelastic scattering
with a leading neutron in the target hemisphere, $ep\to en+X$.  In the
one-pion exchange approximation, this gives access to inclusive deep
inelastic scattering $\gamma^* \pi\to X$ and thus provides information of
the parton densities of the pion.  There is a number of theoretical
investigations focusing on very high energies, as achieved in the HERA
collider experiments \cite{Adloff:1998yg}.  In particular, the studies
\cite{Kopeliovich:1996iw,Holtmann:1996ac,Khoze:2006hw} have investigated
the role of $\rho$ and also of $a_2$ exchange in the framework of Regge
theory.  Furthermore, rescattering of the neutron has been studied in
\cite{Nikolaev:1997cn,D'Alesio:1998bf,Kaidalov:2006cw,Khoze:2006hw} and is
typically found to be non-negligible even for $Q^2$ of several $\gev^2$.
Given the high-energy limit underlying these investigations, we find it
difficult to assess the situation for deeply virtual Compton scattering at
significantly lower energies.

In summary, we find that existing theoretical investigations of similar
processes cannot readily be used to quantify effects beyond the one-pion
exchange approximation for $ep\to e\gamma\pi n$.  They emphasize, however,
the importance of taking $|t|$ as small as possible.  Working at low
$x_\pi$ will in addition help to avoid resonance effects between the
outgoing neutron and pion.  The incorporation of $\rho$ exchange into the
theoretical analysis should be practically feasible (at least at the level
of estimates) if one could gain some information on the size of the
$\rho\to \pi$ transition GPDs, for instance from lattice calculations.

%%%%%%%%%%%%%%%%%%%%%%%%%%%%%%%%%%

\subsection{Compton scattering on the pion}
\label{sec:pi-compton}

Let us now recall the basics of the process $e\pi\to e\gamma\pi$ in the
Bjorken limit, which have been elaborated in detail in earlier work
\cite{Belitsky:2000vk}.  The analysis of this reaction proceeds in close
analogy to the well-known case of a proton target \cite{Diehl:1997bu}.  It
is in fact simpler because the pion has spin zero and thus involves fewer
GPDs and form factors than the proton.  Throughout this section, we retain
only the leading terms in the $1/Q$ expansion, unless explicitly indicated.

We decompose the amplitude for $e\pi\to e\gamma\pi$ into contributions
from Compton scattering and from the Bethe-Heitler process,
\begin{equation}
\mathcal{M}_{e\pi}
 = \mathcal{M}_{\text{VCS}} + \mathcal{M}_{\text{BH}} \,.
\end{equation}
The corresponding decomposition of the differential cross section for
$e\pi\to e\gamma\pi$ and thus also for $ep\to e\gamma\pi n$ reads
\begin{equation}
d\sigma = d\sigma_{\text{VCS}} + d\sigma_{\text{BH}}
        + d\sigma_{\text{INT}} \,,
\end{equation}
where $d\sigma_{\text{INT}}$ is the interference term between the
Bethe-Heitler and Compton processes.
In the Bjorken limit we have for the squared Compton amplitude
\begin{equation}
  \label{vcs}
\sum_{\text{spin}} |\mathcal{M}_{\text{VCS}}|^2
= \frac{4 e^6}{Q^2}\,
  \frac{1-y_\pi^{}+y_\pi^2/2}{y_\pi^2}\,
  |\mathcal{H}_\pi|^2 \,,
\end{equation}
with
\begin{equation}
\mathcal{H}_\pi(\xi,t_\pi) = \sum_q e_q^2 \int_{-1}^1 dx\,
  H_\pi^q(x,\xi,t_\pi)\,
  \biggl[ \frac{1}{\xi-x-i\varepsilon}
        - \frac{1}{\xi+x-i\varepsilon} \biggr]
\end{equation}
at leading order in $\alpha_s$.  Here $H_\pi^q$ is the GPD for quark
flavor $q$ in a $\pi^+$ as defined in \cite{Diehl:2003ny}, $e_q$ is the
quark charge in units of the positron charge $e$, and $\sum_{\text{spin}}$
sums over the polarization of the final-state photon.  The skewness
variable is given by
\begin{equation}
\xi = \frac{\xb^\pi}{2-\xb^\pi} \,.
\end{equation}
The squared Bethe-Heitler amplitude can be written as
\begin{equation}
  \label{bh}
\sum_{\text{spin}} |\mathcal{M}_{\text{BH}}|^2
= \frac{16 e^6}{P}\,
  \frac{1-\xb^\pi}{(\xb^\pi)^2}\, \frac{t_{\pi 0}-t_\pi}{t_\pi^2}\,
  \frac{1-y_\pi^{}+y_\pi^2/2}{1-y_\pi}\,
  \bigl[ F_\pi(t_\pi) \bigr]^2 \,,
\end{equation}
where the factor
\begin{equation}
  \label{prop-fact}
P = \frac{-s' u'}{Q^4\ms (1-y_\pi^{}) /y_\pi^2}
  =  \frac{(A_s - B \cos\phi_\pi) (A_u - B \cos\phi_\pi)}{%
     Q^4\ms (1-y_\pi^{}) /y_\pi^2}
\end{equation}
with $s'= (l'+q')^2$ and $u'= (l-q')^2$ comes from the lepton propagators.
This factor is unity in the Bjorken limit, but it can deviate quite
significantly in experimentally relevant kinematics.  Up to relative
corrections of order $\xb^\pi\ms m^2/Q^2$, one has
\begin{align}
  \label{A-and-B}
A_s &= \frac{Q^2 - (1-y_\pi) t_\pi}{y_\pi} \,,
&
A_u &= \frac{(1-y_\pi) Q^2 - t_\pi}{y_\pi} \,,
&
B   &= \frac{2Q}{y_\pi}\,
             \sqrt{(1-y_\pi) (1-\xb^\pi) (t_{\pi 0}-t_\pi)} \,,
\end{align}
so that for $4 (1-\xb^\pi) (t_{\pi 0}-t_\pi) \sim (1-y_\pi) Q^2$ one can
have $P$ close to zero.  In our numerical calculations we use the exact
expressions of $s'$ and $u'$ in \eqref{prop-fact}.
The interference term between Compton scattering and the Bethe-Heitler
process reads
\begin{align}
  \label{interf}
\sum_{\text{spin}} \mathcal{M}_{\text{VCS}}^*\ms
    \mathcal{M}_{\text{BH}}^{} + \text{c.c.}
 &= e_\ell\, \frac{16 e^6}{P}\,
    \frac{\sqrt{1-\xb^\pi}}{\xb^\pi}\,
    \frac{\sqrt{t_{\pi 0}^{\phantom{2}}-t_\pi^{}}}{Q\ms t_\pi}\,
F_\pi(t_\pi)
\nonumber \\
 &\quad \times
    \biggl[ \frac{1-y_\pi^{}+y_\pi^2/2}{y_\pi \sqrt{1-y_\pi}}\,
            \cos\phi_\pi \re \mathcal{H}_\pi
          + P_\ell\, \frac{1-y_\pi/2}{\sqrt{1-y_\pi}}\,
            \sin\phi_\pi \im \mathcal{H}_\pi \biggr] \,,
\end{align}
where $e_\ell=\pm 1$ is the charge of the lepton beam and $P_\ell=\pm 1$
its helicity.

For lack of better knowledge, we will ignore the off-shellness of the
incoming pion when evaluating $F_\pi(t_\pi)$ and $H_\pi^q(x,\xi,t_\pi)$.
In kinematical factors we take, however, the virtuality $t$ of the initial
pion instead of $m_\pi^2$.  As can be seen in \eqref{tpi-0}, this has an
important effect on $t_{\pi 0}$ and thus on the factors $t_{\pi 0}-t_\pi$
in \eqref{bh} and \eqref{interf}.  The approximation \eqref{rough-approx},
where the $t$-dependence of $d\sigma(e\pi\to e\gamma\pi)$ is neglected,
should therefore be used with caution, especially for small $|t_\pi|$.

In the Bjorken limit, the Bethe-Heitler process dominates over Compton
scattering (unless $y_\pi$ is very small).  This is because
$|\mathcal{M}_{\text{VCS}}|^2 \big/ |\mathcal{M}_{\text{BH}}|^2 \sim t_\pi
/Q^2$ according to \eqref{vcs} and \eqref{bh}.  In this situation,
privileged access to the Compton amplitude is provided by the interference
term \eqref{interf}, which can be separated from the cross section by
reversing the beam charge $e_\ell$ or the beam helicity $P_\ell$.  We
remark that there are also $P_\ell$ dependent terms in
$d\sigma_{\text{VCS}}$.  As can be seen in \cite{Belitsky:2000vk} they
are, however, suppressed by $1/Q$ compared with the dominant term given in
\eqref{vcs}.

To conclude this section we remark on the process $en\to e\gamma\pi^- p$,
which is accessible through incoherent scattering on nuclear targets.
Comparing the subprocesses $e\pi^+\to e\gamma\pi^+$ and $e\pi^-\to
e\gamma\pi^-$, we find that the amplitudes for the Bethe-Heitler process
have opposite sign, whereas those for the Compton process are equal.  This
is because the $\gamma\pi\pi$ three-point function changes sign under
charge conjugation while the $\gamma\pi\pi\gamma$ four-point function
remains the same, and therefore holds even beyond the leading
approximation in $1/Q$.  As a consequence, the relations
\begin{align}
d\sigma_{\text{VCS}}(en\to e\gamma\pi^- p)
 &= \msp d\sigma_{\text{VCS}}(ep\to e\gamma\pi^+ n) \,,
\nonumber \\
d\sigma_{\text{BH}}(en\to e\gamma\pi^- p)
 &= \msp d\sigma_{\text{BH}}(ep\to e\gamma\pi^+ n) \,,
\nonumber \\
d\sigma_{\text{INT}}(en\to e\gamma\pi^- p)
 &= - d\sigma_{\text{INT}}(ep\to e\gamma\pi^+ n)
  \label{pi+pi-}
\end{align}
hold as long as the one-pion exchange approximation is valid, whereas they
will be invalid if for instance interference between $\pi$ and $\rho$
exchange is important.  In typical fixed-target kinematics
$d\sigma_{\text{BH}}$ is much larger than $d\sigma_{\text{INT}}$, so that
the relations \eqref{pi+pi-} can principle be tested experimentally: when
going from $ep\to e\gamma\pi^+ n$ to $en\to e\gamma\pi^- p$, the overall
cross section should approximately remain the same, whereas the beam spin
or beam charge asymmetry should change sign.  We note that a corresponding
consistency check for the one-pion exchange approximation in $ep\to e\pi^+
n$ and $en\to e\pi^- p$ was performed in the extraction \cite{Horn:2006tm}
of the pion form factor.

%%%%%%%%%%%%%%%%%%%%%%%%%%%%%%%%%%%%%%%%%%%%%%%%%%%%%%%%%%%%%

\section{The generalized quark distribution of the pion}
\label{sec:pi-gpd}

In this section we describe the model for the pion GPDs which we will use
in our numerical studies.  We generate a dependence on the skewness $\xi$
by the model of Musatov and Radyushkin \cite{Radyushkin:1998es},
\begin{align}
  \label{dd-models}
H_\pi^q(x,\xi,t_\pi) &=
\int_{-1}^1 d\beta \int_{-1+|\beta|}^{1-|\beta|} d\alpha\;
  \delta(x-\beta-\xi\alpha)\,
  h_b(\beta,\alpha)\, H_\pi^q(\beta,0,t_\pi)
\end{align}
with
\begin{align}
  \label{profile}
h_b(\beta,\alpha) &= \frac{\Gamma(2b+2)}{2^{2b+1}\Gamma^2(b+1)}\,
\frac{[ (1-|\beta|)^2- \alpha^2 ]^b}{(1-|\beta|)^{2b+1}} \,.
\end{align}
For the profile parameter we will take either $b=2$ or $b=1$.  The forward
limit of the GPDs is given by the parton densities in the pion as
$H_\pi^q(x,0,0) = q_\pi(x)$ for $x>0$ and $H_\pi^q(x,0,0) =
-\bar{q}_\pi(-x)$ for $x<0$.  As an input we will take the
parameterizations of SMRS \cite{Sutton:1991ay} or of GRS
\cite{Gluck:1999xe}, both at scale $\mu= 2\gev$.

\begin{figure}
\begin{center}
\psfrag{xx}[c]{$-t_\pi\, [\gev^2]$}
\psfrag{zz}[c]{$F_\pi(t_\pi)$}
\includegraphics[width=0.52\textwidth]{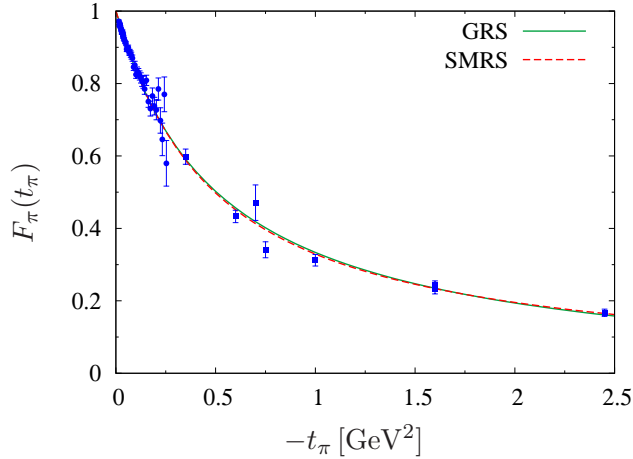}
\caption{\label{fig:pi-ff} Data for the electromagnetic pion form factor
  from \protect\cite{Amendolia:1986wj,Ackermann:1977rp,%
    Tadevosyan:2007yd,Horn:2006tm}, compared to a fit specified by
  \protect\eqref{DFJK4-ansatz}, \protect\eqref{DFJK4-profile}, and the
  parameters in \protect\eqref{ABfit}.}
\end{center}
\end{figure}

\begin{figure}
\begin{center}
\psfrag{xx}[c]{$x$}
\psfrag{zz}[c]{$b\, [\mathrm{fm}]$}
\includegraphics[width=0.52\textwidth]{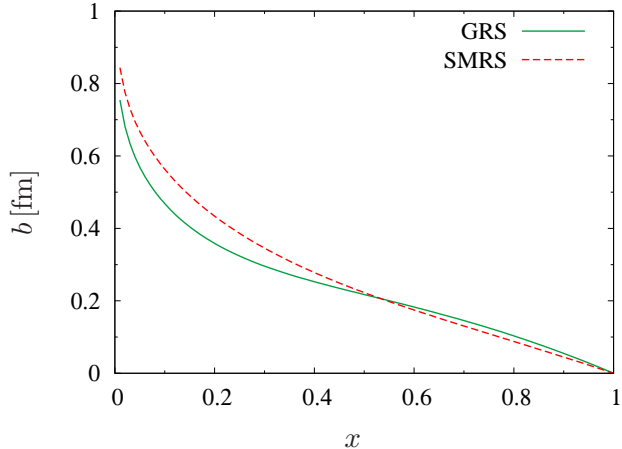}
\caption{\label{fig:pi-impact} The average transverse distance
  $\sqrt{\langle b^2 \rangle_x}$ between a $u$-quark and the center of
  momentum of the $\pi^+$, as obtained in the two fits of
  Fig.~\ref{fig:pi-ff}.}
\end{center}
\end{figure}

The simplest way to model the $t_\pi$ dependence is a factorized ansatz
\begin{equation}
  \label{factorized}
H_\pi^q(x,0,t_\pi) = H_\pi^q(x,0,0)\ms F_\pi(t_\pi) \,,
\end{equation}
which automatically satisfies the sum rule
\begin{equation}
  \label{sum-rule}
\sum_q e_q \int_{-1}^1 dx\, H_\pi^q(x,0,t_\pi) = F_\pi(t_\pi) \,.
\end{equation}
Both theoretical considerations \cite{Burkardt:2004bv} and lattice QCD
calculations \cite{Brommel:2005ee,Brommel:2007zz} indicate, however, that
the dependence of the GPDs on $t_\pi$ and $x$ is correlated.  As a model
ansatz we will take an exponential $t_\pi$ dependence with an $x$
dependent slope.  Such an ansatz has proven to be quite successful for the
proton \cite{Diehl:2004cx,Guidal:2004nd}.  Following \cite{Diehl:2004cx}
we set
\begin{equation}
  \label{DFJK4-ansatz}
H_\pi^q(x,0,t_\pi) = H_\pi^q(x,0,0)\ms \exp\bigl[ t_\pi f(|x|) \bigr]
\end{equation}
with
\begin{equation}
  \label{DFJK4-profile}
f(x) = \alpha' (1-x)^3 \ln\frac{1}{x}
           + B\ms (1-x)^3 + A\ms x (1-x)^2 \,,
\end{equation}
where $\alpha' = 0.9\gev^{-2}$ is taken in accordance with Regge
phenomenology.  We fit $A$ and $B$ to describe the pion form factor
through the sum rule \eqref{sum-rule}, using the data
\cite{Amendolia:1986wj,Ackermann:1977rp,Tadevosyan:2007yd,Horn:2006tm} as
selected in \cite{Brommel:2006ww}.  With the fitted values
\begin{align}
  \label{ABfit}
A &= 2.19 \gev^{-2} \,, & B &= -0.38 \gev^{-2} &
  & \text{for GRS,}
\nonumber \\
A &= 1.35 \gev^{-2} \,, & B &= \msp 0.58 \gev^{-2}  &
  & \text{for SMRS}
\end{align}
we get a good description for $F_\pi(t_\pi)$ with both parameterizations
of the parton densities, as shown in Fig.~\ref{fig:pi-ff}.  An intuitive
physical interpretation of the function $f(x)$ is obtained in the impact
parameter representation \cite{Burkardt:2002hr}: for $x>0$ the average of
the squared transverse distance between the quark and the center of
momentum of the pion is $\langle b^2 \rangle_x = 4 f(x)$.  The results of
our two fits are rather similar and yield physically reasonable values, as
shown in Fig.~\ref{fig:pi-impact}.

As a caveat we note that the sum rule \eqref{sum-rule} only constrains the
valence quark distributions, given by $H^q(x,0,t_\pi) + H^q(-x,0,t_\pi) =
\bigl[ q_\pi(x) - \bar{q}_\pi(x) \bigr] \exp\bigl[ t_\pi f(x) \bigr]$ for
$x>0$, and is insensitive to the sea quarks.  Since sea quarks mix with
gluons under evolution, one may expect that the $t_\pi$ dependence in this
sector is different from the one for valence distributions.  The ansatz
\eqref{DFJK4-ansatz} with a common $t_\pi$ dependence for valence and sea
quarks (including the strange sea) may hence be regarded as
oversimplified.  Given, however, that even in the forward limit the sea
quark distributions in the pion are poorly known at present, we deem this
ansatz to be acceptable for our purpose.

\begin{figure}
\begin{center}
\psfrag{xx}[c]{$-t_\pi\, [\gev^2]$}
\psfrag{zz}[c]{$A_{20}^u(t_\pi)$}
\includegraphics[width=0.5\textwidth]{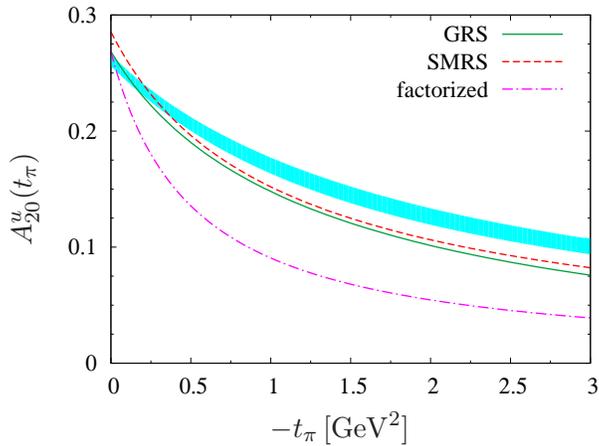}
\caption{\label{fig:pi-mom} The second moment \protect\eqref{A20} of the
  pion GPD $H_\pi^u$, obtained with the two fits of Fig.~\ref{fig:pi-ff}.
  Also shown is the result of the factorized ansatz
  \protect\eqref{factorized} with the GRS parton densities.  The band
  corresponds to a parameterization of lattice results given in
  \protect\cite{Brommel:2007zz}, $A_{20}^u(t_\pi) = A_{20}^u(0)
  \big/\bigl(1 - t_\pi/M^2\bigr)$ with $A_{20}^{u}(0) = 0.261(5)$ and
  $M=1.37(7) \gev$.}
\end{center}
\end{figure}

Using \eqref{DFJK4-ansatz} and \eqref{DFJK4-profile} we can also evaluate
the second moment
\begin{equation}
  \label{A20}
A_{20}^u(t_\pi) = \int_{-1}^1 dx\, x H_\pi^u(x,0,t_\pi) \,,
\end{equation}
which has been evaluated in lattice QCD
\cite{Brommel:2005ee,Brommel:2007zz}.  In Fig.~\ref{fig:pi-mom} we compare
the results of our fits \eqref{ABfit} with a monopole parameterization of
the lattice data given in \cite{Brommel:2007zz}.  Although not perfect,
the agreement is quite good, and certainly much better than the result of
the factorized ansatz \eqref{factorized}.  We note that, in our
parameterization, the contribution of sea quarks to the moment \eqref{A20}
is below $30\%$ at $t_\pi=0$ and smaller at nonzero $t_\pi$.

Notwithstanding the success of the ansatz given by \eqref{DFJK4-ansatz}
and \eqref{DFJK4-profile} in reproducing the lowest two moments of the
pion GPD, some cautionary remarks from the theoretical side are in order.
As discussed in \cite{Diehl:2004cx}, the asymptotic behavior of the pion
form factor at large negative $t$ in our ansatz is controlled by the
Feynman mechanism and given by the Drell-Yan relation $F_\pi(t) \sim
|t|^{-(1+\beta)/2}$.  Here $\beta$ describes the behavior $q(x) \sim
(1-x)^\beta$ of the parton densities in the limit $x\to 1$ and is
predicted to be $\beta=2$ for the pion \cite{Lepage:1980fj,Yuan:2003fs}.
Asymptotically the Feynman mechanism hence gives a contribution $F_\pi(t)
\sim |t|^{-3/2}$ to the form factor, which is power suppressed compared
with the contribution $F_\pi(t) \sim |t|^{-1}$ from the hard-scattering
mechanism \cite{Efremov:1979qk,Lepage:1980fj}, where it is understood that
both power laws are modified by logarithms.  The ansatz
\eqref{DFJK4-ansatz}, \eqref{DFJK4-profile} does not have the form
generated by the hard-scattering mechanism for $H_\pi^q(x,\xi,t)$ at large
$t$ \cite{Vogt:2001if}.  However, it turns out that the parton density
parameterizations \cite{Gluck:1999xe,Sutton:1991ay} at $\mu= 2\gev$ both
have a large-$x$ power $\beta\approx 1$ within less than $10\%$, in stark
contrast with the prediction $\beta=2$ from power counting.  Thus, our
ansatz gives $F_\pi(t) \sim |t|^{-1}$ at large negative $t$, which is
compatible with the monopole behavior that describes very well the
available data \cite{Brommel:2006ww}.  We shall not pursue this issue
further, and use our ansatz for $H_\pi^q(x,\xi,t_\pi)$ as a simple
candidate form that is not in contradiction with phenomenological
constraints.

%%%%%%%%%%%%%%%%%%%%%%%%%%%%%%%%%%%%%%%%%%%%%%%%%%%%%%%%%%%%%

\section{Cross section estimates}
\label{sec:cross-est}

In this section, we calculate the cross section for $ep\to en\gamma\pi$
and its dependence on the beam charge and beam helicity.  Since the
unpolarized cross section is dominated by the Bethe-Heitler process, the
corresponding results are largely model independent (as long as the
one-pion exchange approximation is adequate).  On the other hand, we will
see that the charge and polarization asymmetries for the lepton beam are
quite sensitive to the model we assume for the pion GPD.  We will compare
four simple model scenarios:
\begin{enumerate}
\item a skewness dependence generated by \eqref{dd-models} and
  \eqref{profile} with $b=2$ and a $t_\pi$ dependence given by
  \eqref{DFJK4-ansatz} to \eqref{ABfit}, with the GRS parton densities in
  the pion,
\item the same as model 1, but with the SMRS parton densities,
\item the same as model 1, but with $b=1$ instead of $b=2$,
\item the same as model 1, but with a factorizing $t_\pi$ dependence
  \eqref{factorized}.  We use this model for the sake of contrast,
  although it is disfavored by theoretical considerations and lattice
  data.
\end{enumerate}
For the pion form factor we use a monopole parameterization $F_\pi(t_\pi)
= 1 \big/\bigl( 1 - t_\pi/M^2 \bigr)$ with $M= 714 \mev$.  This provides a
very good description of the experimental data, as shown in
\cite{Brommel:2006ww}.

As explained in Sect.~\ref{sec:kin}, we impose minimal values for $Q^2$
and $s_\pi$ and at the same time a maximal value for $|t|$, which requires
a sufficiently large energy.  We therefore concentrate on typical
kinematics for HERMES and for the planned Jefferson Lab upgrade to $11
\gev$ beam energy.  The leading-twist interpretation of DVCS demands that
$|t_\pi| \ll Q^2$, so that we also put a cut $|t_\pi| <
|t_\pi|_{\text{max}}$.
We finally impose a maximum value on $y$ (and thus on $y_\pi$).  On the
experimental side, this ensures that the scattered lepton has sufficient
energy to be detected and identified.  On the theoretical side, this
improves the $1/Q$ expansion underlying the approximate formulae
\eqref{bh} and \eqref{interf}, since there are subleading terms in $1/Q$
that come with a factor $1/\sqrt{1-y_\pi}$ relative to the leading terms.
An example of such a subleading term is found in the propagator factor
$P$, see \eqref{prop-fact} and \eqref{A-and-B}.

Let us first consider the case of HERMES, with a beam energy $E_e = 27.6
\gev$ in the proton rest frame.  We impose a lower limit $\vartheta_\gamma
> 2.57^\circ$ on the angle between the momenta of the final-state photon
and the lepton beam in the target rest frame.  This value corresponds to
the maximal geometric acceptance of the electromagnetic calorimeter in the
experiment, see e.g.\ Sect.~5.22 of \cite{Ye:2006pe}.
Taking limiting values
\begin{align}
  \label{def-cuts}
Q_{\text{min}}^2 &= 2 \gev^2 \,,
&
s_{\pi\ms\text{min}} &= 4 \gev^2 \,,
&
y_{\text{max}} &= 0.85
\end{align}
and
\begin{align}
  \label{big-t}
\tmax &= 0.5\gev^2 \,,
&
|t_\pi|_{\text{max}} &= 0.9\gev^2 \,,
\end{align}
we find a Bethe-Heitler cross section of $\sigma_{\text{BH}} = 1620 \fb$,
which is between $15\%$ and $20\%$ smaller than the result we obtain in
model 1 for $\sigma_{\text{BH}} + \sigma_{\text{VCS}} +
\sigma_{\text{INT}}$ with either an electron or positron beam.  With an
integrated luminosity of order $1 \fb^{-1}$ for HERMES running on a proton
target \cite{Stewart:2007pc}, we deem this cross section to be too small,
since it will be further decreased by experimental acceptance cuts and
detection efficiency, and since according to \eqref{interf} the extraction
of the beam charge or beam polarization asymmetry requires a differential
measurement at least in the angle $\phi_\pi$.  Loosening the requirements
\eqref{def-cuts} or \eqref{big-t} would increase the rate at the price of
going to kinematics where the theoretical interpretation used in this
paper becomes increasingly questionable.

\begin{figure}
\begin{center}
\psfrag{xx}[c]{$E_\pi\, [\gev]$}
\psfrag{zz}[c]{$\vartheta_\pi\, [\text{deg}]$}
\includegraphics[width=0.48\textwidth]{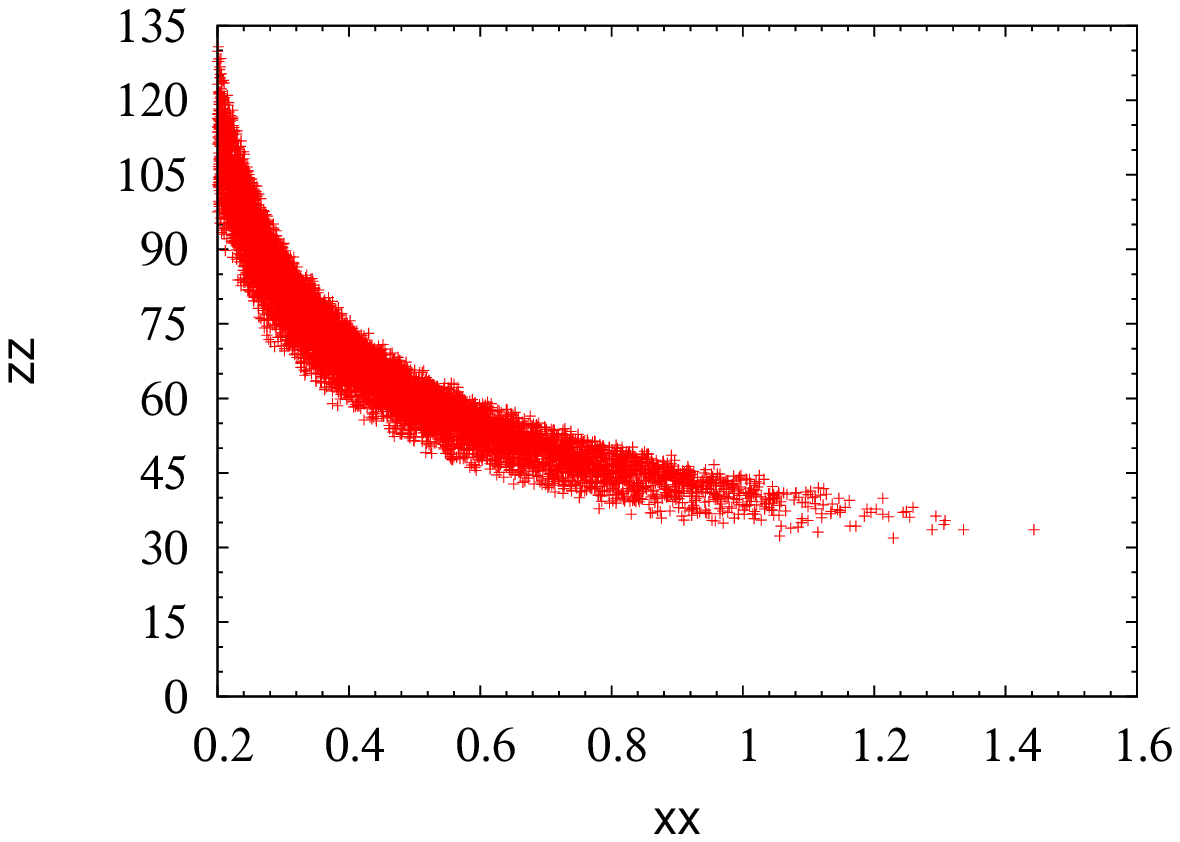}
\hspace{1em}
\psfrag{xx}[c]{$E_\gamma\, [\gev]$}
\psfrag{zz}[c]{$\vartheta_\gamma\, [\text{deg}]$}
\includegraphics[width=0.48\textwidth]{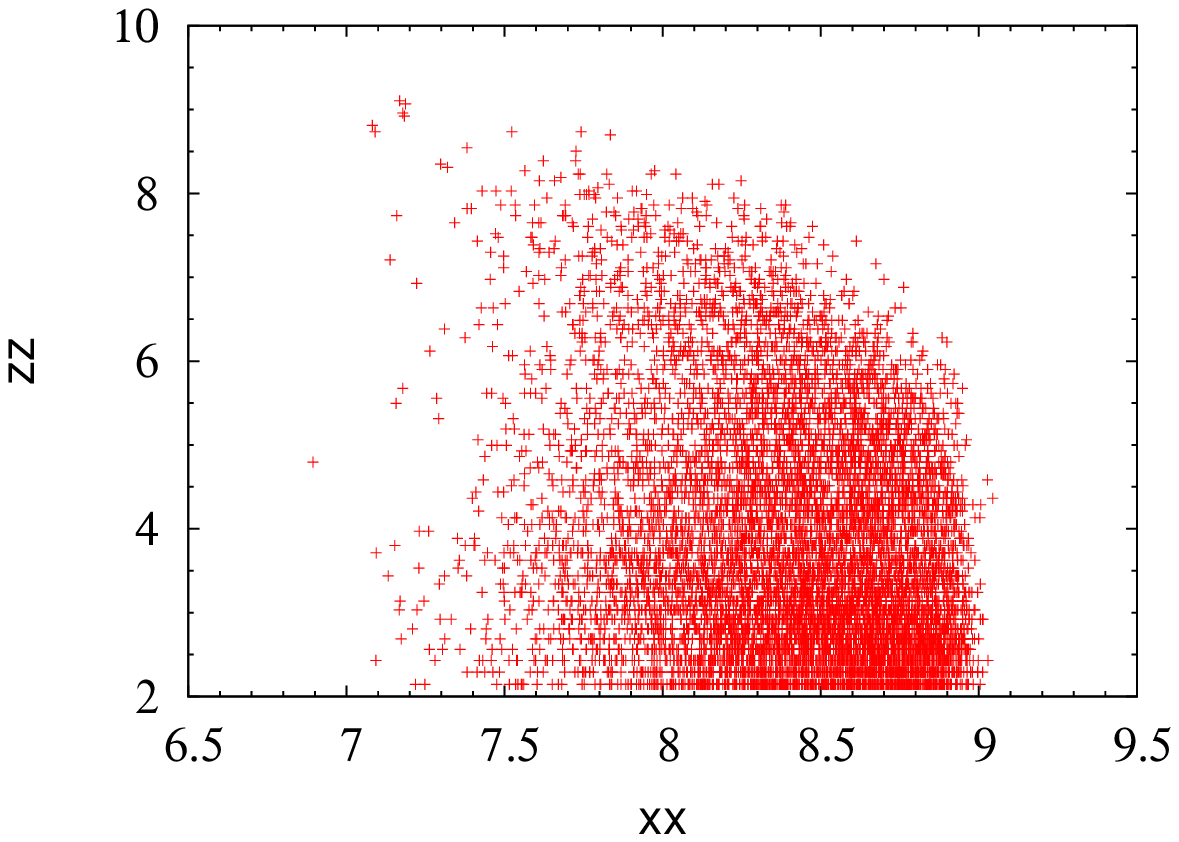}
\caption{\label{fig:scat-plots} Scatter plots for the energies and polar
  angles of the final-state pion and photon in the proton rest frame.  The
  plots are generated from the Bethe-Heitler cross section within the
  kinematics specified by \protect\eqref{def-cuts},
  \protect\eqref{exp-cuts}, and \protect\eqref{small-t}.}
\end{center}
\end{figure}

Higher luminosities than at HERMES can be achieved by the experiments at
Jefferson Lab.  With a currently available beam energy of up to $E_e =
6\gev$, the requirements \eqref{def-cuts} and \eqref{big-t} leave no
available phase space, as can be seen from the bounds on $x_\pi$ in
\eqref{xpi-max} and \eqref{xpi-y-min}.  This will be changed with the
energy upgrade to $E_e = 11 \gev$, which we consider in the remainder of
this section.  We assume that the outgoing electron, photon, and pion are
detected experimentally.  Note that identification of the pion (or of the
recoiling neutron) is necessary to distinguish the signal process $ep\to
e\gamma\pi^+ n$ from DVCS on the proton, $ep\to e\gamma p$, which has a
far greater rate.
In the proton rest frame we have
\begin{align}
  \label{Epi-lab}
E_\pi &= \frac{1}{2 \mn^{} x_\pi}
 \biggl\{ m_\pi^2-t_\pi^{}
      - \bigl[ (1- x_\pi^{})(1-\xb^\pi) - \xb^\pi \ms\bigr]\ms t
\nonumber \\
 &\qquad\qquad\quad
  + 2 \cos(\phi_\pi +\psi_e-\psi_n)\,
    \sqrt{(1-x_\pi^{}) (1-\xb^\pi) (t_0-t) (t_{\pi 0}-t_\pi)} \,\biggr\}
 + \mathcal{O}\Bigl( \frac{m^2}{Q} \Bigr) \,,
\nonumber \\
\cos\vartheta_\pi &=
   \biggl( 1 - \frac{x_\pi^{} (1-\xb^\pi)\ms \mn^{}}{E_\pi} \biggr)
   \frac{E_\pi}{\sqrt{E_\pi^2 - m_\pi^2}}
 + \mathcal{O}\Bigl( \frac{m}{Q} \Bigr) \,,
\end{align}
where $E_\pi$ is the energy of the final-state pion and $\vartheta_\pi$
the angle between its momentum and the lepton beam direction.  Likewise,
we find
\begin{align}
E_\gamma &= \frac{Q^2}{2 \xb\ms \mn} + \mathcal{O}(m) \,,
&
\cos\vartheta_\gamma
 &= 1 - \frac{2 (1-y)\ms \xb^2 \mn^2}{Q^2}
  + \mathcal{O}\Bigl( \frac{m^3}{Q^3} \Bigr)
\end{align}
for the energy of the outgoing photon and its polar angle.  To estimate
the cross section and its dependence on the beam charge and polarization,
we assume some minimal experimental cuts, which correspond to the
acceptance planned for the CLAS$++$ detector \cite{Guidal:2007pc},
\begin{align}
  \label{exp-cuts}
E'_e    &> 500 \mev \,,
&
8^\circ &< \vartheta_e < 45^\circ \,,
\nonumber \\
E_\gamma &> 100 \mev \,,
&
2^\circ & < \vartheta_\gamma < 40^\circ \,,
\nonumber \\
E_\pi   &> 200 \mev \,,
&
5^\circ &< \vartheta_\pi < 135^\circ \,.
\end{align}
Typical values of these quantities are shown in the scatter plots of
Fig.~\ref{fig:scat-plots}.  These plots have been generated using the
Bethe-Heitler cross section, which dominates the total rate as we shall
see shortly.  We have imposed the kinematic requirements \eqref{def-cuts}
and
\begin{align}
  \label{small-t}
\tmax &= 0.3\gev^2 \,,
&
|t_\pi|_{\text{max}} &= 0.7\gev^2 \,,
\end{align}
where compared with \eqref{big-t} we have taken smaller $\tmax$ and
$|t_\pi|_{\text{max}}$, so that the one-pion exchange and the
leading-twist approximations are better fulfilled.  We see that the pion
has small to intermediate energy and covers a large angular region,
whereas the photon is energetic and strongly focused in the beam
direction.  In the corresponding scatter plot for the outgoing electron
kinematics (not shown here) we have $1.6\gev < E'_e < 3.4\gev$ and
$13^\circ < \vartheta_e < 25^\circ$, so that the experimental cuts on
these quantities in \eqref{exp-cuts} have no influence.

\begin{figure}
\begin{center}
\psfrag{xx}[c]{$\xb^\pi$}
\psfrag{zz}[c]{$Q^2\, [\gev^2]$}
\includegraphics[width=0.48\textwidth]{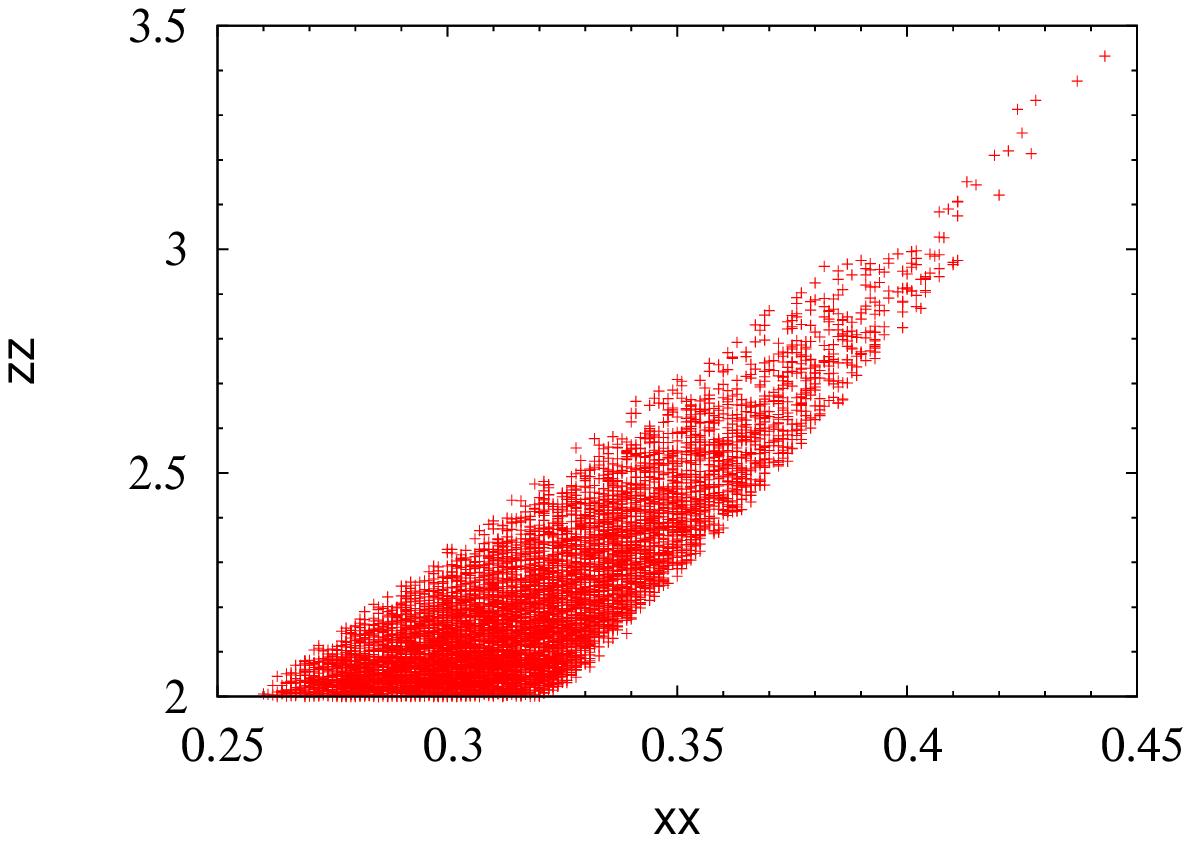}
\hspace{1em}
\psfrag{xx}[c]{$\xb^\pi$}
\psfrag{zz}[c]{$s_{\pi}\, [\gev^2]$}
\includegraphics[width=0.48\textwidth]{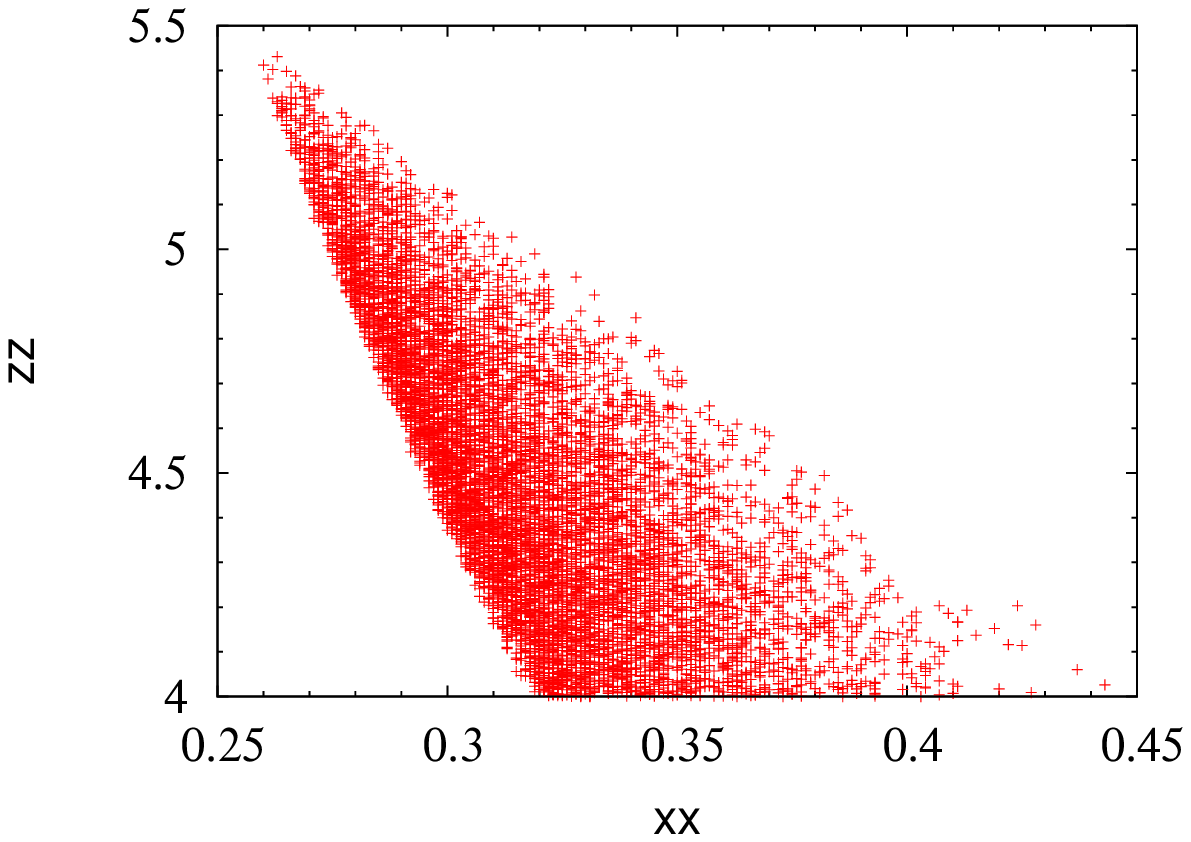}
\\[2em]
\psfrag{xx}[c]{$-t_\pi\, [\gev^2]$}
\psfrag{zz}[c]{$M^2_{\pi n}\, [\gev^2]$}
\includegraphics[width=0.48\textwidth]{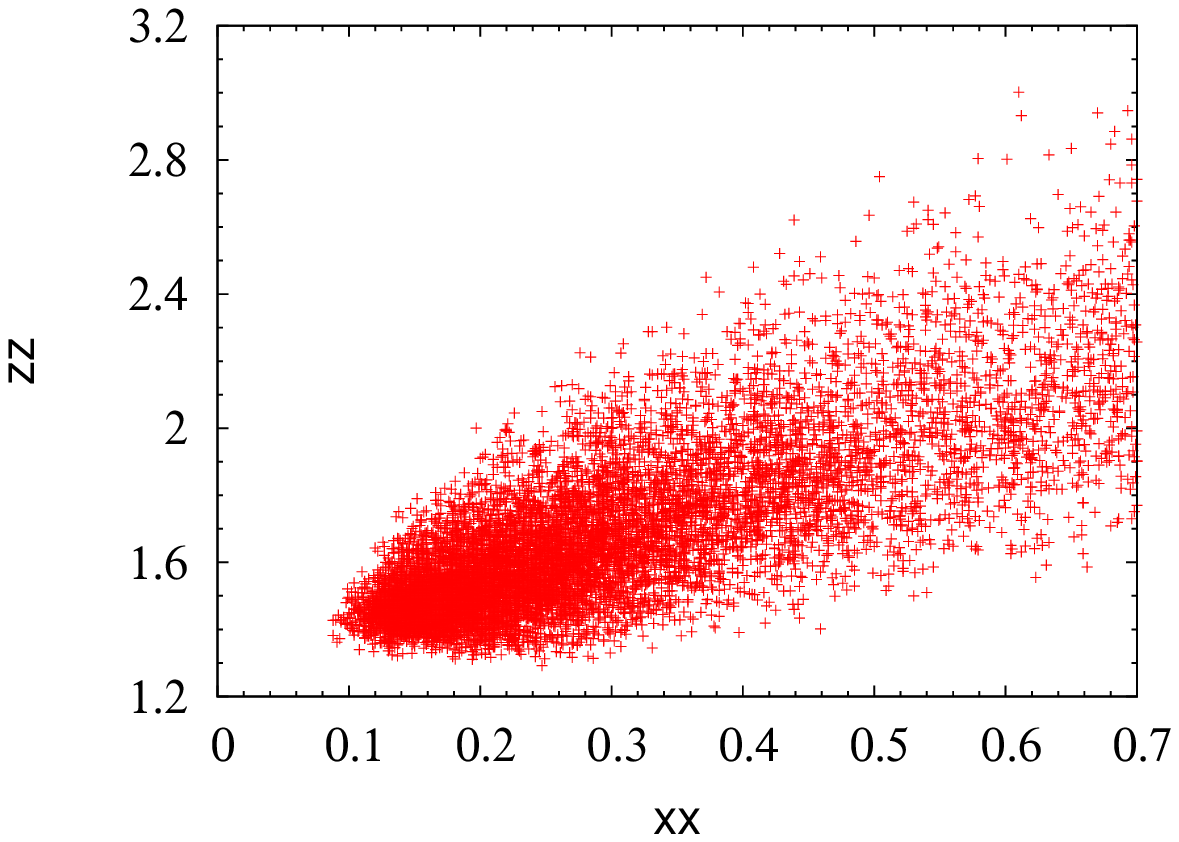}
\caption{\label{fig:s-plots} Scatter plots for various kinematic
  quantities, generated in the same way as in
  Fig.~\protect\ref{fig:scat-plots}.}
\end{center}
\end{figure}

Figure~\ref{fig:s-plots} shows corresponding scatter plots for other
relevant variable pairs.  In the first two panels we see the values of
$Q^2$, $s_\pi$, and $\xb^\pi$ at which the Compton process $\gamma^*\pi\to
\gamma\pi$ can be probed with $E_e=11 \gev$.  The last panel shows that
typical values of the squared $\pi n$ invariant mass $M^2_{\pi n}$ are
between $1.3 \gev^2$ and $2.2 \gev^2$, which unfortunately includes the
region of nucleon resonances.  We will come back to this point shortly.

In addition to the contributions $\sigma_{\text{BH}}$,
$\sigma_{\text{VCS}}$, $\sigma_{\text{INT}}$ of the different subprocesses
and their interference to the integrated cross section, we evaluate the
weighted differences
\begin{align}
  \label{weights}
S_{C}^{\cos\phi_\pi} &= \int d\phi_\pi\, \cos\phi_\pi\,
  \biggl[ \frac{d\sigma(e_\ell=+1)}{d\phi_\pi}
        - \frac{d\sigma(e_\ell=-1)}{d\phi_\pi} \biggr] \,,
\nonumber \\
S_{L}^{\sin\phi_\pi} &= \int d\phi_\pi\, \sin\phi_\pi\,
  \biggl[ \frac{d\sigma(P_\ell=+1)}{d\phi_\pi}
        - \frac{d\sigma(P_\ell=-1)}{d\phi_\pi} \biggr] \,,
\end{align}
of cross sections for different beam charge or beam polarization.  In the
approximation given by \eqref{vcs}, \eqref{bh}, \eqref{interf}, they are
respectively proportional to $\re \mathcal{H}_\pi$ and $\im
\mathcal{H}_\pi$ in the interference term.  In the Bjorken limit, the
propagator factor $P$ in \eqref{prop-fact} becomes $\phi_\pi$ independent,
so that $\smash{S_C^{\cos\phi_\pi}}$ and $\smash{S_L^{\sin\phi_\pi}}$
respectively give the coefficients of $\cos\phi_\pi$ and $\sin\phi_\pi$ in
$2\pi\ms d\sigma_{\text{INT}}/d\phi_\pi$.  With $E_e=11 \gev$ and the
kinematics delineated by \eqref{def-cuts} and \eqref{small-t}, we find,
however, a clear $\phi_\pi$ dependence for $P$.  For this reason, the
interference term \eqref{interf} also contributes to the cross section
integrated over $\phi_\pi$, as seen in Table~\ref{tab:cuts}.  We also
recall that beyond the leading approximation in $1/Q$, the weighted cross
section $\smash{S_L^{\sin\phi_\pi}}$ receives a contribution from
$d\sigma_{\text{VCS}}$ in addition to the one from $d\sigma_{\text{INT}}$.
This can be seen from the expressions in \cite{Belitsky:2000vk} and is
well-known in the case of DVCS on a proton \cite{Diehl:1997bu}.

% version in color
\begin{table}
  \caption{\label{tab:cuts} The contributions of the Bethe-Heitler and
    Compton processes and of their interference to the integrated cross
    section, as well as the weighted cross sections defined in
    \protect\eqref{weights}.  Results are evaluated in model 1 for different
    kinematic constraints in addition to the cuts \protect\eqref{exp-cuts}.
    The signs of $\sigma_{\text{INT}}$ and $\smash{S_L^{\sin\phi_\pi}}$ refer
    to an electron beam ($e_\ell= -1$).  Limiting values of $Q^2$,
    $s_\pi$, $t$, $t_\pi$, and $M^2_{\pi n}$ are given in units of $\gev^2$,
    and cross sections in units of $\fb$.}
\begin{center}
\renewcommand{\arraystretch}{1.2}
\begin{tabular}{lccclclllll} \hline
$Q^{2}_{\text{min}}$ & $s_{\pi\ms\text{min}}$
 & $\tmax$ & $|t_\pi|_{\max}$ & $y_{\text{max}}$
 & $M^2_{\pi n\, \text{min}}$
 & $\sigma^{}_{\text{BH}}$ & $\sigma^{}_{\text{VCS}}$
 & $\sigma^{}_{\text{INT}}$
 & $S_C^{\cos\phi_\pi}$ & $S_L^{\sin\phi_\pi}$ \\
\hline
$2$   & $4$ & $0.3$ & $0.7$ & $0.85$ & ---
 & $\phantom{1}18.4$ & $0.88$ & $-0.18$ & $0.39$ & $\phantom{1}7.57$ \\
$2$   & $4$ & $0.3$ & $0.7$ & \colorbox{yellow}{$0.8$} & ---
 & $\phantom{11}5.12$ & $0.29$ & $-0.09$ & $0.17$ & $\phantom{1}2.17$ \\
$2$   & $4$ & $0.3$ & $0.7$ & \colorbox{yellow}{$0.9$} & ---
 & $\phantom{1}45.6$ & $1.86$ & $-0.27$ & $0.64$ & $17.9$ \\
$2$   & $4$ & \colorbox{yellow}{$0.2$} & $0.7$ & $0.85$ & ---
 & $\phantom{10}0.41$ & $0.016$ & $-0.002$ & $0.004$
 & $\phantom{1}0.16$ \\
$2$   & $4$ & \colorbox{yellow}{$0.5$} & $0.7$ & $0.85$ & ---
 & $105$ & $6.52$ & $-2.32$ & $5.00$ & $46.2$ \\
\colorbox{yellow}{$2.5$} & $4$ & $0.3$ & $0.7$ & $0.85$ & ---
 & $\phantom{10}2.55$ & $0.103$ & $-0.010$ & $0.018$
 & $\phantom{1}0.96$ \\
$2$   & \colorbox{yellow}{$5$} & $0.3$ & $0.7$ & $0.85$ & ---
 & $\phantom{10}0.30$ & $0.013$ & $-0.003$ & $0.008$
 & $\phantom{1}0.12$ \\
$2$   & $4$ & $0.3$ & \colorbox{yellow}{$0.5$} & $0.85$ & ---
 & $\phantom{1}16.2$ & $0.69$ & $-0.09$ & $0.18$ & $\phantom{1}6.30$ \\
$2$   & $4$ & $0.3$ & $0.7$ & $0.85$ & \colorbox{yellow}{$1.5$}
 & $\phantom{1}13.4$ & $0.67$ & $-0.19$ & $0.42$
 & $\phantom{1}5.72$ \\
$2$   & $4$ & $0.3$ & $0.7$ & $0.85$ & \colorbox{yellow}{$1.8$}
 & $\phantom{10}5.08$ & $0.31$ & $-0.14$ & $0.30$
 & $\phantom{1}2.46$ \\
\hline
\end{tabular}
\end{center}
\end{table}

In Table~\ref{tab:cuts} we show our results for the integrated and
weighted cross sections calculated in model 1 for different kinematic
constraints.  For the choice in \eqref{def-cuts} and \eqref{small-t},
shown in the first row, one obtains an integrated cross section of $18.4
\fb$.  With an estimated luminosity of $3000 \fb^{-1}$ per year at
CLAS$++$, this gives a very comfortable rate of about $55000$ events, so
that one may hope that even with realistic experimental cuts and detection
efficiencies there will be sufficient statistics to perform a binning in
several variables.  

The further entries in Table~\ref{tab:cuts} illustrate how the situation
changes if one modifies our baseline kinematic constraints.  Raising
$y_{\text{max}}$ from $0.85$ to $0.9$ would more than double the rate, but
as discussed above this would make the $1/Q$ expansion underlying the
formulae \eqref{bh} and \eqref{interf} worse.  Lowering instead
$y_{\text{max}}$ from $0.85$ to $0.8$, would reduce the rate by a factor
of about $3.5$.  Increasing $\tmax$ from $0.3\gev^2$ to $0.5\gev^2$ we
obtain a significantly higher rate, but the one-pion exchange
approximation is much more problematic in that case.  Decreasing $\tmax$
to $0.2 \gev^2$ would improve the quality of the one-pion exchange
approximation, but the resulting loss of rate is probably too much for
such a cut to be useful.  Taking a stricter cut of $2.5 \gev^2$ instead of
$2 \gev^2$ for $Q^{2}_{\text{min}}$ would make the leading-twist analysis
of DVCS safer but decrease the cross section by about a factor of 7.  An
even stronger decrease is found if one requires $s_\pi$ to be above
$5\gev^2$ instead of $4\gev^2$.  By contrast, only very little rate is
lost if one takes $0.5\gev^2$ instead of $0.7\gev^2$ for
$|t_{\pi}|_{\text{max}}$, so that it may be worthwhile to consider a
stronger cut on this variable.  Finally, one may wish to impose a cut on
the invariant mass of the $\pi n$ system, so as to reduce possible
resonance effects in this channel, which are not taken into account in our
theoretical description.  The entries in the last two rows of
Table~\ref{tab:cuts} respectively correspond to a lower cut on $M_{\pi n}$
at the mass of the $\Delta$ and at the mass of the $\Delta$ plus its
width.  The resulting rates show that at least part of the kinematic
region where resonance effects may spoil a simple interpretation can be
removed in an analysis without losing too much signal.

\begin{figure}
\begin{center}
\psfrag{xx}[c]{$-t\, [\gev^2]$}
\psfrag{zz}[c]{$d\sigma_{\text{BH}}/dt\; [\fb /\gev^{2}]$}
\includegraphics[width=0.48\textwidth]{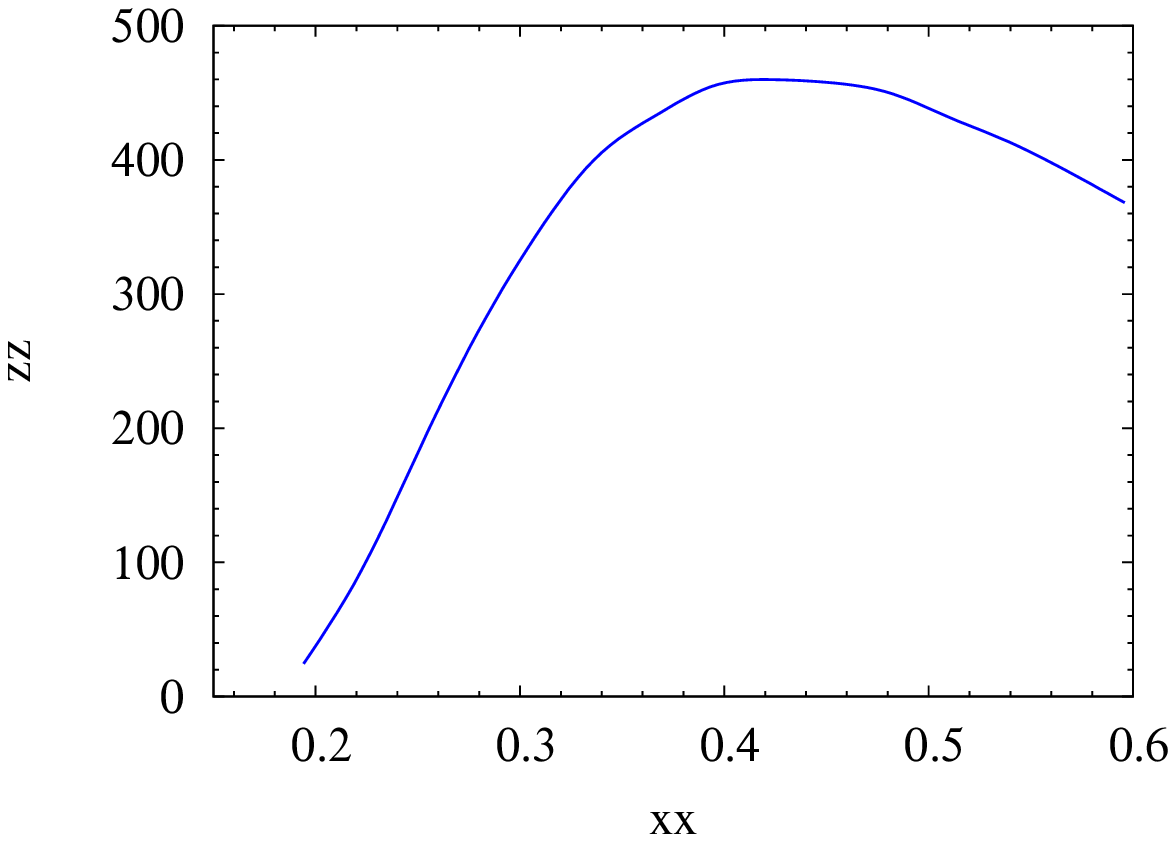}
\hspace{1em}
\psfrag{xx}[c]{$-t_\pi\, [\gev^2]$}
\psfrag{zz}[c]{$d\sigma_{\text{BH}}/dt_\pi\; [\fb /\gev^{2}]$}
\includegraphics[width=0.48\textwidth]{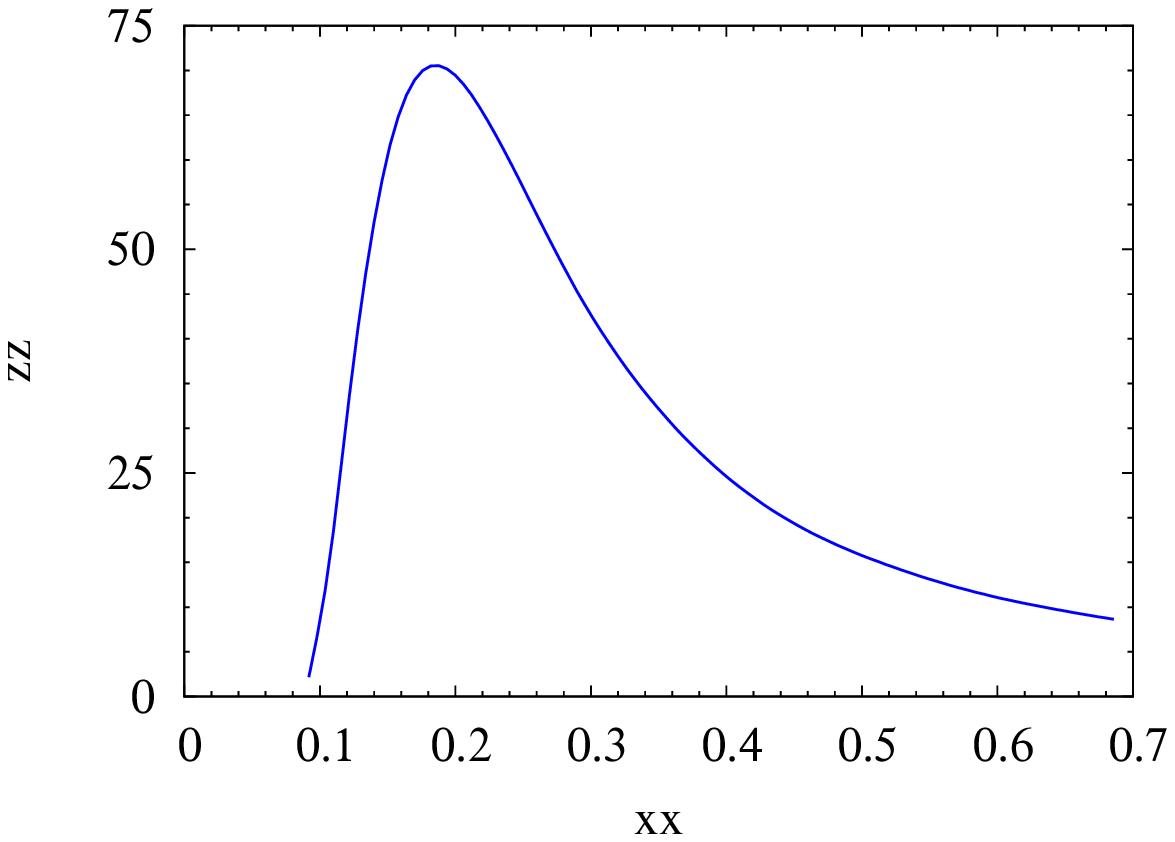}
\\[2em]
\psfrag{xx}[c]{$x_\pi$}
\psfrag{zz}[c]{$d\sigma_{\text{BH}}/dx_\pi\; [\fb]$}
\includegraphics[width=0.48\textwidth]{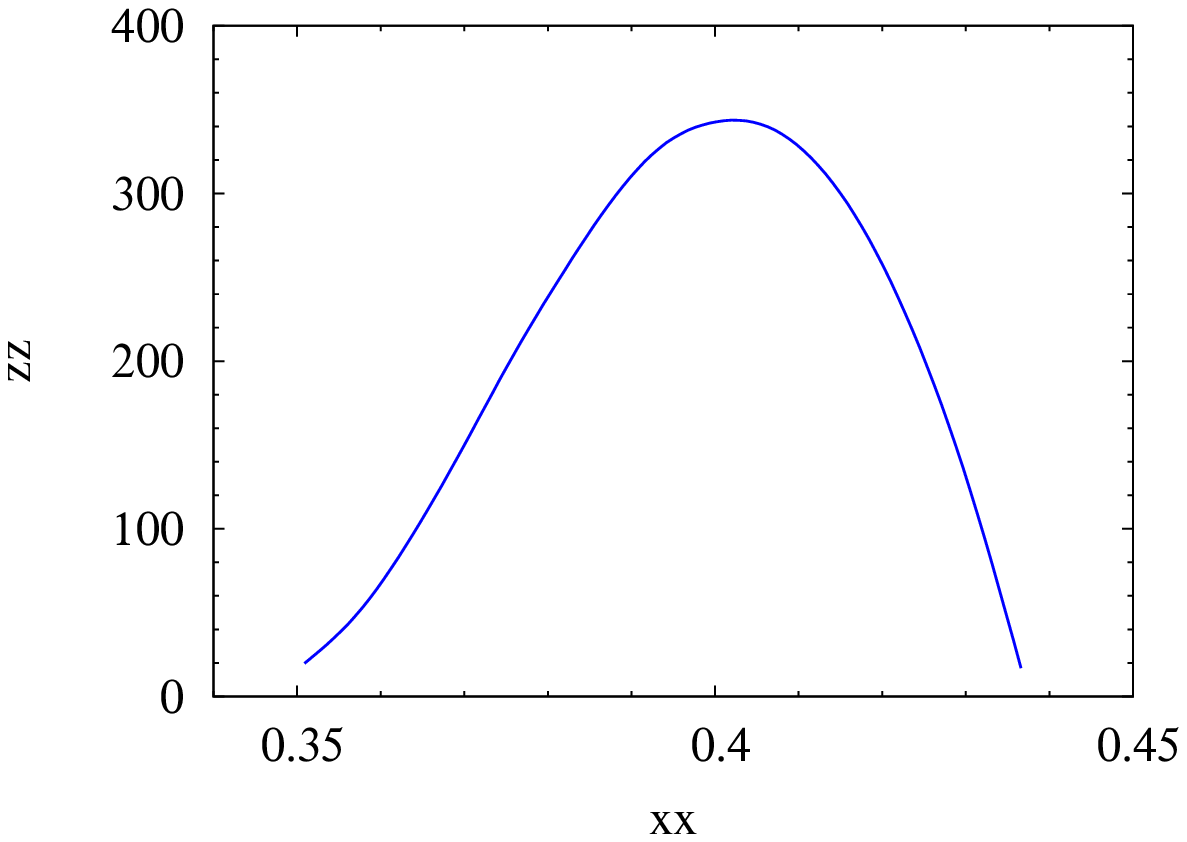}
\caption{\label{fig:diff-bh} The Bethe-Heitler cross section taken
  differential in $t$, $t_\pi$, or $x_\pi$, and calculated with the cuts
  specified by \protect\eqref{def-cuts}, \protect\eqref{exp-cuts}, and
  \protect\eqref{small-t}.  For the cross section differential in $t$ we
  have omitted the first constraint in \protect\eqref{small-t}.}
\end{center}
\end{figure}

In Fig.~\ref{fig:diff-bh} we show the Bethe-Heitler cross section
differential in $t$, $t_\pi$, or in $x_\pi$.  We see that in the
kinematics of Jefferson Lab at $11 \gev$ a cutoff $\tmax$ much below $0.3
\gev^2$ severely restricts the available phase space.  The $t_\pi$
spectrum falls off much more slowly than the corresponding $t$ spectrum
for DVCS on the proton, which readily follows from the slower decrease of
the pion electromagnetic form factor compared with the proton form
factors.  Nevertheless, the falloff for $|t_\pi| \gsim 0.2\gev^2$ is
sufficiently steep to account for the weak dependence of the cross section
on $|t_\pi|_{\text{max}}$ seen in Table~\ref{tab:cuts}.  The spectrum in
$x_\pi$ is rather featureless and reflects the form of the pion flux
factor in Fig.~\ref{fig:pi-flux} together with the phase space boundaries
$x_{\text{min}}$ and $x_{\text{max}}$ given in \eqref{xpi-max} and
\eqref{xpi-y-min}.

\begin{table}
\begin{center}
\caption{\label{tab:models} As Table~\protect\ref{tab:cuts} but for
  different models of the pion GPDs.  The kinematic constraints in
  \protect\eqref{def-cuts}, \protect\eqref{exp-cuts}, and
  \protect\eqref{small-t} are always assumed.  The corresponding
  Bethe-Heitler cross sections is $\sigma^{}_{\text{BH}}= 18.4 \fb$.}
\vspace{0.5em}
\renewcommand{\arraystretch}{1.2}
\begin{tabular}{crrrrr} \hline
model & $\sigma^{}_{\text{VCS}}$ & $\sigma^{}_{\text{INT}}$
 & $S_C^{\cos\phi_\pi}$ & $S_L^{\sin\phi_\pi}$ \\
\hline
1 & $0.88$ & $-0.18$ & $0.39$ & $7.57$ \\
2 & $0.67$ & $-0.66$ & $1.57$ & $6.44$ \\
3 & $1.12$ & $-0.21$ & $0.43$ & $8.57$ \\
4 & $0.70$ & $0.25$ & $-0.62$ & $6.78$ \\
\hline
\end{tabular}
\end{center}
\end{table}

\begin{figure}
\begin{center}
\psfrag{xx}[c]{$-t\, [\gev^2]$}
\psfrag{zz}[c][t]{$dS_C^{\cos\phi_\pi}/dt\; [\fb /\gev^{2}]$}
\includegraphics[width=0.48\textwidth]{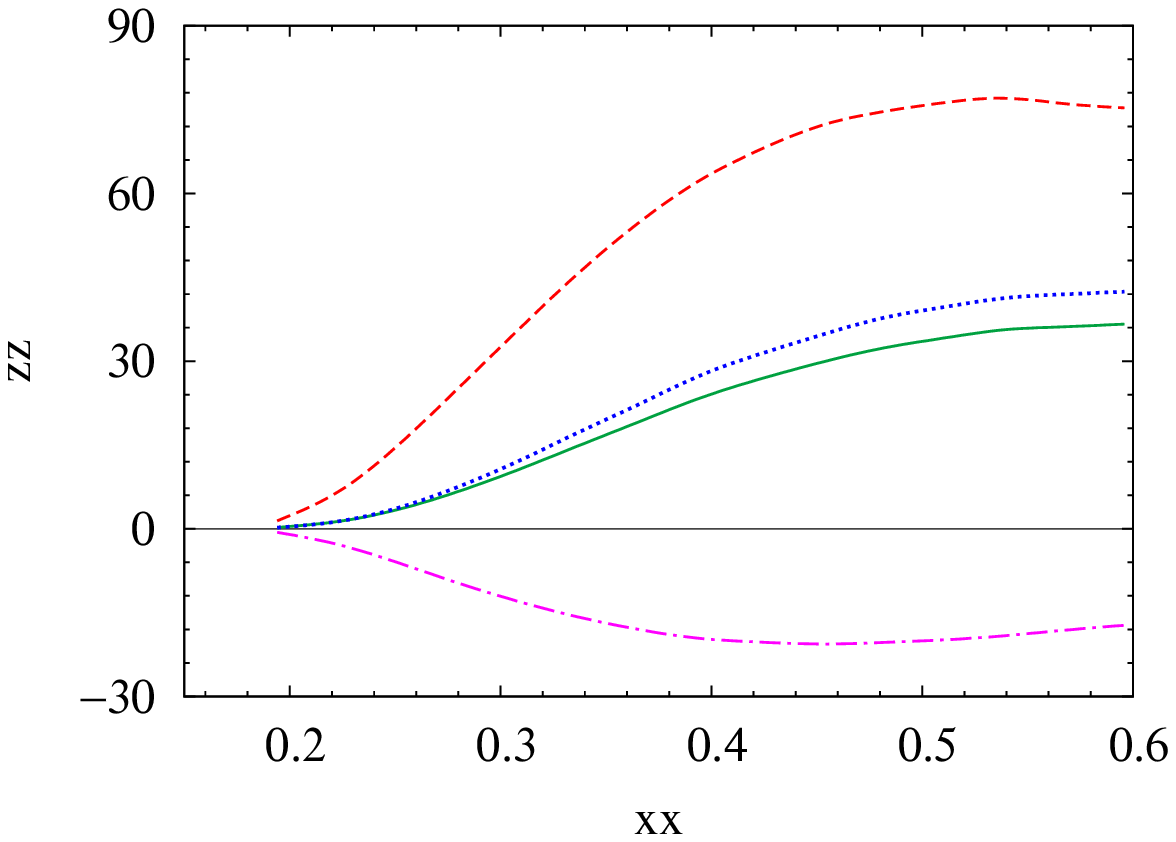}
\hspace{1em}
\psfrag{zz}[c][t]{$dS_L^{\sin\phi_\pi}/dt\; [\fb /\gev^{2}]$}
\includegraphics[width=0.48\textwidth]{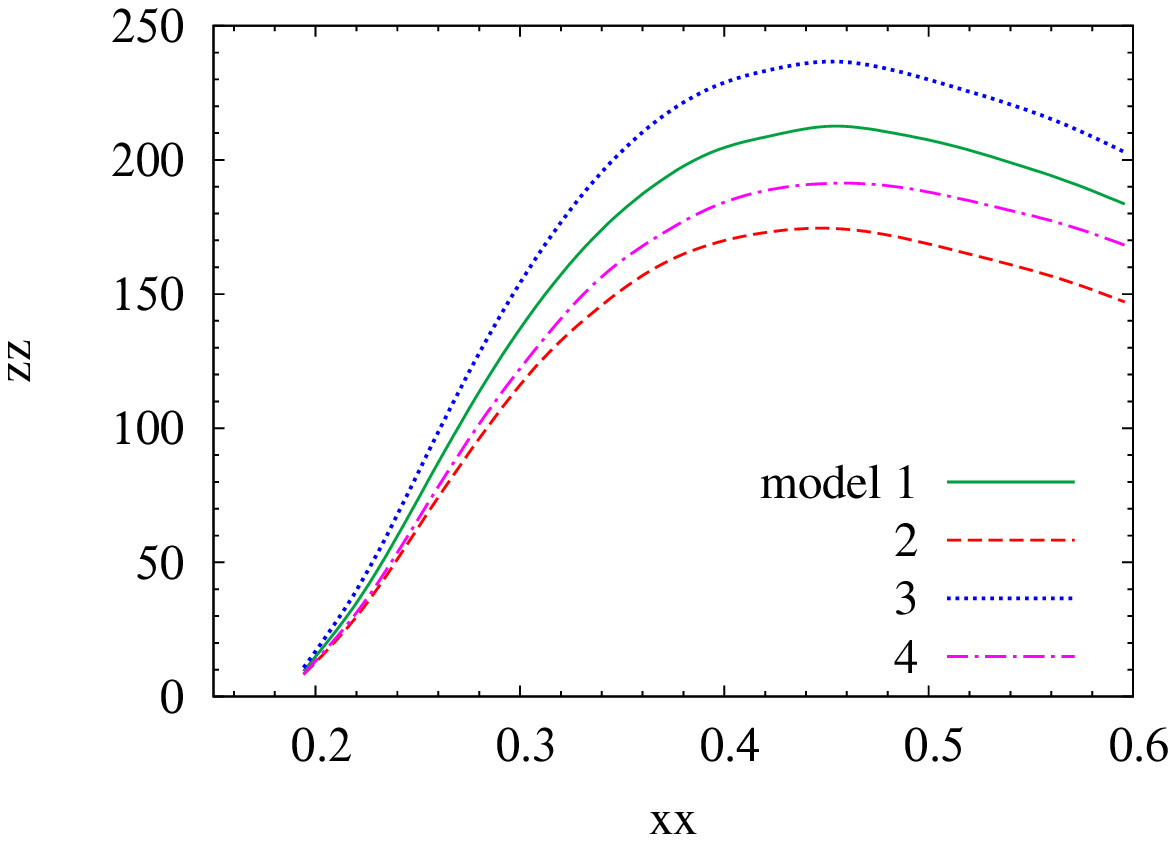}
\\[1.5em]
\psfrag{xx}[c]{$-t_\pi\, [\gev^2]$}
\psfrag{zz}[c][t]{$dS_C^{\cos\phi_\pi}/dt_\pi\; [\fb /\gev^{2}]$}
\includegraphics[width=0.48\textwidth]{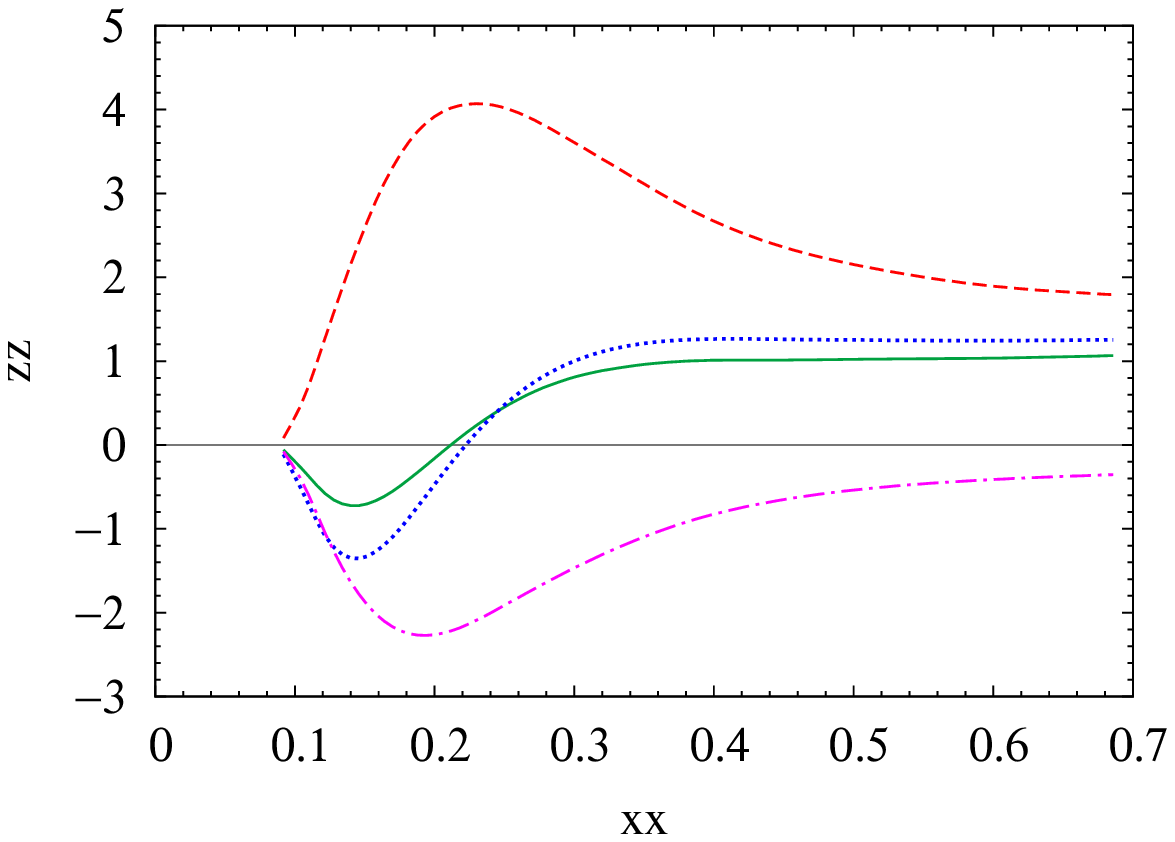}
\hspace{1em}
\psfrag{zz}[c][t]{$dS_L^{\sin\phi_\pi}/dt_\pi\; [\fb /\gev^{2}]$}
\includegraphics[width=0.48\textwidth]{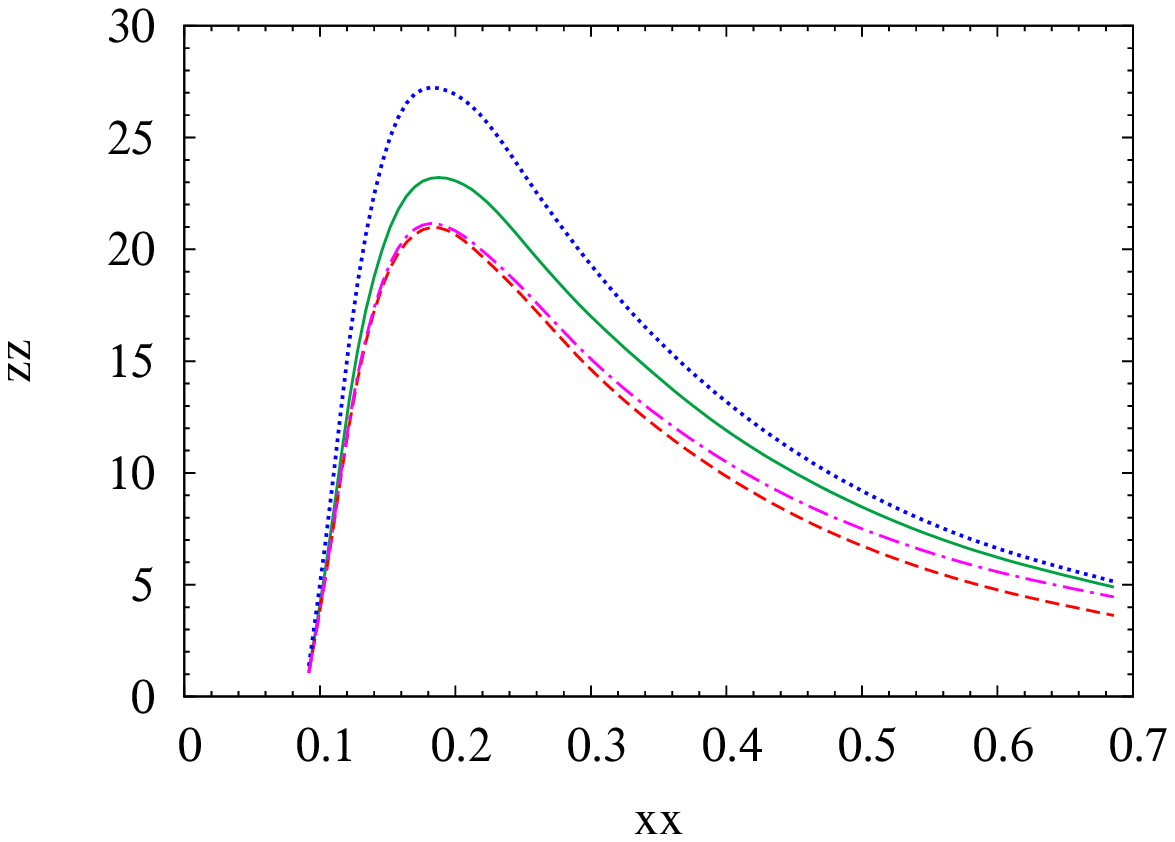}
\\[1.5em]
\psfrag{xx}[c]{$x_\pi$}
\psfrag{zz}[c][t]{$dS_C^{\cos\phi_\pi}/dx_\pi$}
\includegraphics[width=0.48\textwidth]{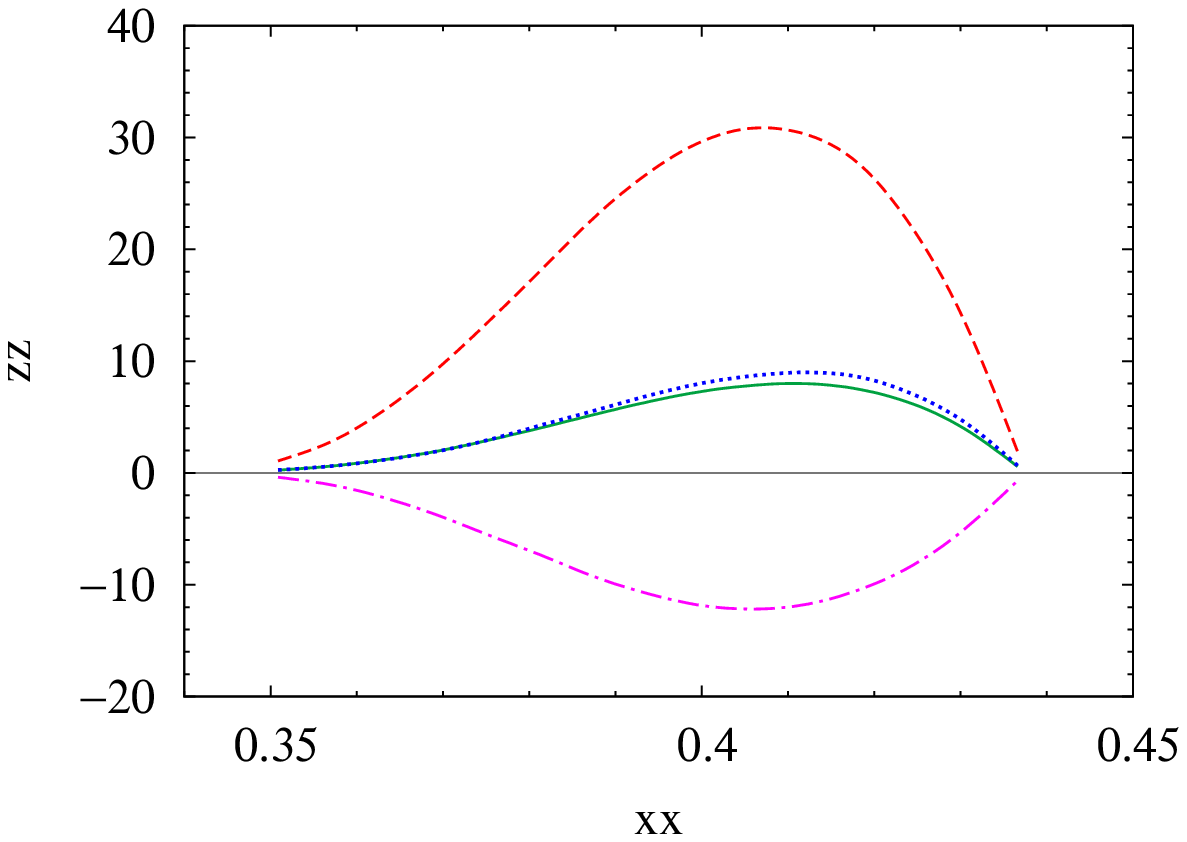}
\hspace{1em}
\psfrag{zz}[c][t]{$dS_L^{\sin\phi_\pi}/dx_\pi$}
\includegraphics[width=0.48\textwidth]{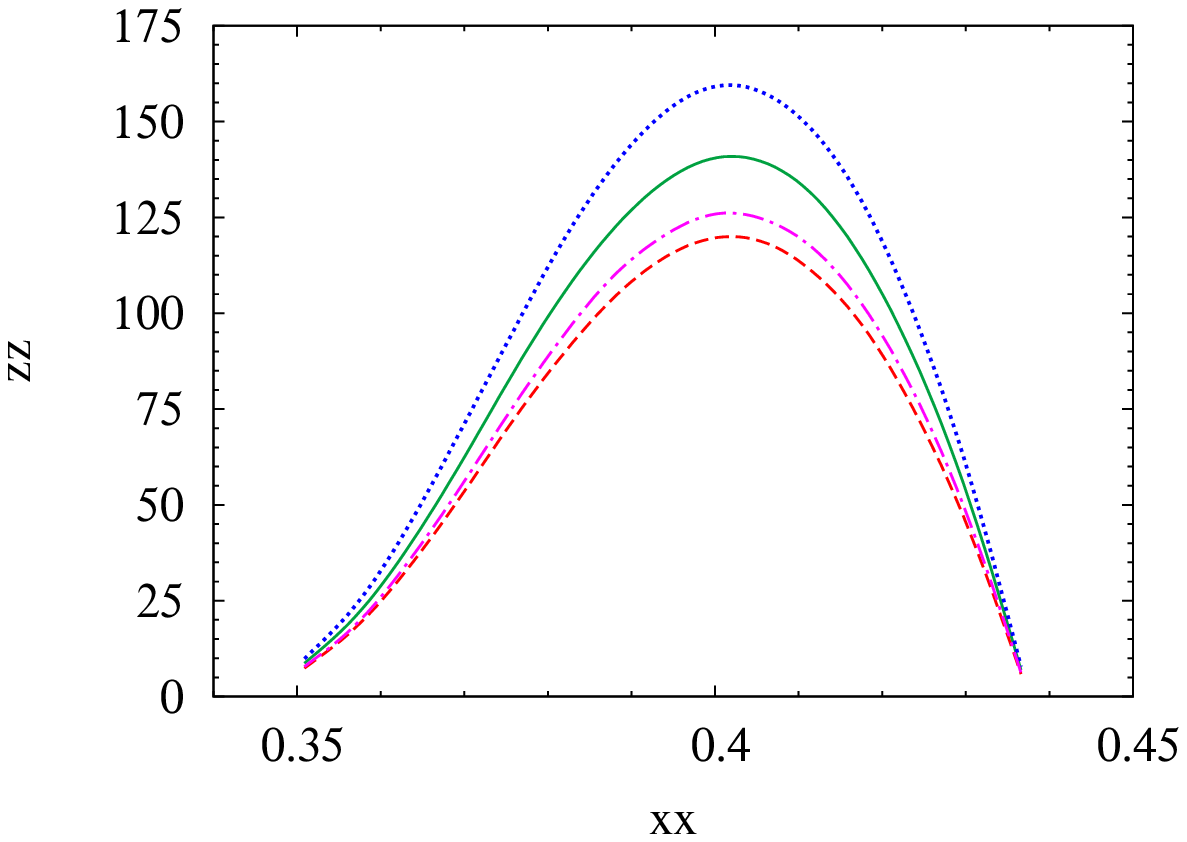}
\caption{\label{fig:diff-weight} As Fig.~\protect\ref{fig:diff-bh} but for
 the weighted cross sections \protect\eqref{weights}, calculated with the
 four GPD models specified in the text.  The sign of $S_L^{\sin\phi_\pi}$
 refers to an electron beam.}
\end{center}
\end{figure}

The sensitivity to the pion GPDs of the Compton cross section and the
interference term is documented in Table~\ref{tab:models} for the
different models introduced above.  To obtain a more detailed picture, we
plot in Fig.~\ref{fig:diff-weight} the weighted cross sections
\eqref{weights} differential in $t$, $t_\pi$, and $x_\pi$.  We see that
the spread between models is much more pronounced for
$\smash{S_C^{\cos\phi_\pi}}$, with different signs and even a zero
crossing in $t_\pi$ for some of the models.  By contrast, the variation of
$S_L^{\sin\phi_\pi}$ is less drastic and concerns the size of the weighted
cross section more than its dependence on $t$, $t_\pi$, or $x_\pi$.  On
the other hand, the beam spin asymmetry $S_L^{\sin\phi_\pi}$ is
significantly larger than the beam charge asymmetry $S_C^{\cos\phi_\pi}$
and may be easier to measure in practice.

%%%%%%%%%%%%%%%%%%%%%%%%%%%%%%%%%%%%%%%%%%%%%%%%%%%%%%%%%%%%%

\section{Summary} 
\label{sec:sum}

We have investigated the possibility to study DVCS on the pion in the
reaction $ep\to e\gamma\pi^+ n$.  Such a study is experimentally demanding
for several reasons.  Firstly, the phase space is limited by the
requirements of small $t$ on one side (so that pion exchange dominates the
process and the pion is not too far off-shell) and of large $Q^2$ and
$s_\pi$ on the other side (so that an analysis based on the Bjorken limit
is applicable).  As can be seen from \eqref{xpi-max} and
\eqref{xpi-y-min}, this favors experiments with a higher $ep$ c.m.\
energy.  Secondly, the cross section for $ep\to e\gamma\pi^+ n$ is
significantly smaller than the one for $ep\to e\gamma p$, which puts high
demands on both the luminosity and the experimental identification of the
final state.

We find that conditions for an experimental study of this process are not
favorable in current experiments, with HERMES being limited by the
available event rate and Jefferson Lab by the beam energy.  After the
planned energy upgrade to $11 \gev$ at Jefferson Lab, it should, however,
be possible to investigate the reaction in detail.  Using acceptance cuts
relevant for the CLAS$++$ detector, we find comfortable event rates for
kinematical conditions that may not be optimal but should be adequate for
a first look at DVCS on the pion in the Bjorken regime and at intermediate
values of the skewness~$\xi$.  Using simple models for the GPDs of the
pion, we estimate that information about them could be provided both by
the beam spin and by the beam charge asymmetry, with the latter showing a
more pronounced sensitivity to the ansatz for the GPDs but being smaller
in size.  Optimized studies might be feasible one day at a projected
electron-ion collider \cite{Deshpande:2005wd}, where in particular it
should be possible to take the invariant mass of the $\pi n$ system above
the resonance region.

As a part of our model study, we have explored an ansatz for the pion GPDs
that depends exponentially on $t_\pi$ with a slope decreasing with $x$.
Taking a functional form previously used for the nucleon
\cite{Diehl:2004cx} together with current parameterizations of the parton
densities in the pion, we obtain an excellent fit to the experimental data
of the electromagnetic pion form factor.  Without adjusting further
parameters, the ansatz then compares rather well with recent results from
lattice QCD for the second Mellin moment of the GPDs.

%%%%%%%%%%%%%%%%%%%%%%%%%%%%%%%%%%%%%%%%%%%%%%%%%%%%%%%%%%%%%
%%%%%%%%%%%%%%%%%%%%%%%%%%%%%%%%%%%%%%%%%%%%%%%%%%%%%%%%%%%%%

\section*{Acknowledgments} 

We gratefully thank H.~Avagyan, A.~Bacchetta, J.~Bartels, D.~Br\"ommel,
M.~Guidal, D.~Hasch, M.~Kopytin, C.~Riedl, W.~Kugler, S.~Niccolai,
B.~Pire, J.~Stewart and L.~Szymanowski for useful discussions and
correspondence.  Special thanks go to B.~Pire for useful remarks on the
manuscript.
This work is supported by the exchange program PROCOPE of the German
Academic Exchange Service and the French Minist\`ere des Affaires
\'Etrang\`eres, by the Integrated Infrastructure Initiative of the
European Union (contract number RII3-CT-2004-506078), and by the Helmholtz
Association (contract number VH-NG-004).  D.A. and J.P.L. thank the
members of the IFPA group at the University of Li\`ege for their
hospitality.

%%%%%%%%%%%%%%%%%%%%%%%%%%%%%%%%%%%%%%%%%%%%%%%%%%%%%%%%%%%%%

\end{document}